\documentclass{aa} 
\usepackage{graphicx}
\usepackage{natbib}
\usepackage{txfonts} 
\newcommand{\eCMa}{$\epsilon$ CMa}
\newcommand{\rgd}{$R_\mathrm{G/D}$}
\newcommand{\mG}{$\mu \mathrm{G}$}

\newcommand{\commentpcf}[1]{}
\newcommand{\kms}{km s$^{-1}$}
\newcommand{\cmtwo}{cm$^{-2}$}
\newcommand{\cc}{cm$^{-3}$}
\newcommand{\psec}{s$^{-1}$}

\newcommand{\NaI}{\ion{Na}{i}}
\newcommand{\CaII}{\ion{Ca}{ii}}
\newcommand{\CII}{\ion{C}{ii}}
\newcommand{\NCII}{$N($\ion{C}{ii}$)$}
\newcommand{\NCIIstar}{$N($\ion{C}{ii}$^*)$}
\newcommand{\CIIstar}{\ion{C}{ii}$^*$}
\newcommand{\NNI}{$N($\ion{N}{i}$)$}
\newcommand{\NNII}{$N($\ion{N}{ii}$)$}

\newcommand{\NMgI}{$N($\ion{Mg}{i}$)$}
\newcommand{\NMgII}{$N($\ion{Mg}{ii}$)$}

\newcommand{\NH}{$N($H$)$}
\newcommand{\ebv}{$E(B-V)$}

\newcommand{\OI}{\ion{O}{i}}

\newcommand{\OVII}{O$^{+7}$}
\newcommand{\OVIII}{O$^{+8}$}
\newcommand{\NI}{\ion{N}{i}}
\newcommand{\NII}{\ion{N}{ii}}

\newcommand{\NeII}{\ion{Ne}{ii}}
\newcommand{\NeIII}{\ion{Ne}{iii}}
\newcommand{\lya}{Ly$\alpha$}
\newcommand{\HI}{\ion{H}{i}}

\newcommand{\SiII}{\ion{Si}{ii}}
\newcommand{\SiIII}{\ion{Si}{iii}}
\newcommand{\MgII}{\ion{Mg}{ii}}
\newcommand{\MgI}{\ion{Mg}{i}}
\newcommand{\HeI}{\ion{He}{i}}
\newcommand{\HII}{\ion{H}{ii}}
\newcommand{\HH}{H$_2$}

\newcommand{\NHI}{$N($\ion{H}{i}$)$}
\newcommand{\NHII}{$N($\ion{H}{ii}$)$}

\newcommand{\nHI}{$n($\ion{H}{i}$)$}
\newcommand{\nHeI}{$n($\ion{He}{i}$)$}

\newcommand{\meanH}{$\langle n_\mathrm{H}\rangle$}
\newcommand{\nel}{$n_\mathrm{e}$}
\newcommand{\np}{$n_\mathrm{p}$}

\newcommand{\HO}{H$^0$}
\newcommand{\HeO}{He$^0$}
\newcommand{\nHO}{$n(\mathrm{H}^0)$}
\newcommand{\nHeO}{$n(\mathrm{He}^0)$}

\begin{document}

\title{The Boundary Conditions of the Heliosphere:  Photoionization Models
Constrained by Interstellar and \emph{In Situ} Data}

\author{Jonathan D. Slavin\inst{1} \and Priscilla C. Frisch \inst{2}}
\institute{Harvard-Smithsonian Center for Astrophysics, 60 Garden St., MS 83,
Cambridge, MA 02138  \\
\email{jslavin@cfa.harvard.edu} 
\and
University of Chicago, 
Department of Astronomy and Astrophysics, 
5640 S.\ Ellis Ave., Chicago, IL 60637 \\
\email{frisch@oddjob.uchicago.edu}}
\offprints{J. Slavin}
\date{Received -- / Accepted --}

\abstract
{The boundary conditions of the heliosphere are set by the ionization,
density and composition of inflowing interstellar matter.}
{Constraining the properties of the Local Interstellar Cloud (LIC) at
the heliosphere requires radiative transfer ionization models.}  
{We model the background interstellar radiation field using observed stellar
FUV and EUV emission and the diffuse soft X-ray background.  We also model the
emission from the boundary between the LIC and the hot Local Bubble (LB)
plasma, assuming that the cloud is evaporating because of thermal conduction.
We create a grid of models covering a plausible range of LIC and LB
properties, and use the modeled radiation field as input to radiative
transfer/thermal equilibrium calculations using the Cloudy code.  Data from
\emph{in situ} observations of \HeO, pickup ions and anomalous cosmic rays
in the heliosphere,
and absorption line measurements towards $\epsilon$~CMa were used to constrain
the input parameters.}
{A restricted range of assumed LIC \ion{H}{i} column densities and LB plasma
temperatures produce models that match all the observational constraints.  The
relative weakness of the constraints on \NHI\ and $T_h$ contrast with the
narrow limits predicted for the \HO\ and electron density in the LIC at the
Sun, $n(\mathrm{H}^0) = 0.19 - 0.20 $ \cc, and \nel\ $=0.07 \pm 0.01$ \cc.
Derived abundances are mostly typical for low density gas, with sub-solar Mg,
Si and Fe, possibly subsolar O and N, and S about solar; however C is
supersolar.}  
{The interstellar gas  at the Sun is warm, low density, and partially ionized, 
with $n(\mathrm{H}) \approx 0.23 - 0.27$ \cc, $T \approx 6300$ K,
$X(\mathrm{H}^+) \sim 0.2$, and $X(\mathrm{He}^+) \sim 0.4$. These
results appear to be robust since acceptable models are found for
substantially different input radiation fields.  Our results favor low values
for the reference solar abundances for the LIC composition.}

\titlerunning{Boundary Conditions of the Heliosphere}
\authorrunning{Slavin and Frisch} 
\maketitle

\section{Introduction} 

The solar wind evacuates a cavity in the interstellar medium (ISM) known as
the heliosphere, from which interstellar ions are excluded.  In contrast,
neutral interstellar gas flows through the heliosphere until destroyed by
charge-transfer with the solar wind and photoionization. These neutrals form
the parent population of pickup ions (PUI) and anomalous cosmic rays (ACR)
observed inside of the heliosphere.  The properties of the surrounding
interstellar medium set the boundary conditions of the heliosphere and
determine its configuration and evolution. An ionization gradient is expected
in the cloud feeding ISM into the heliosphere because of the hardness of the
interstellar radiation field and the low opacity of the ISM \citep[][hereafter
SF02]{Cheng+Bruhweiler_1990,Vallerga:1996,Slavin+Frisch_2002}.  Because of this
ionization gradient, the densities of partially ionized species in the local
interstellar cloud (LIC) will differ from the values in the circumheliosphere
interstellar medium (CHISM) that forms the boundary conditions of the
heliosphere.  Hence we have undertaken a series of studies to determine the
boundary conditions of the heliosphere based on both astronomical and
heliospheric data. In turn our results provide tighter constraints on the
heliosphere models used to calculate the filtration factors for neutrals that
then permit comparisons between ISM inside and outside of the heliosphere.

The distribution and velocity of interstellar \HO\ inside of the heliosphere
were first determined over 30 years ago from the florescence of solar \lya\
radiation from these atoms
\citep{ThomasKrassa:1971,BertauxBlamont:1971,AdamsFrisch:1977}.  Similar
observations of solar 584\AA\ florescence from interstellar \HeO\ showed that
H/He $\sim 6$ for interstellar gas inside of the heliosphere, in contrast to the
cosmic value H/He $=10$ \citep{Ajello:1978,WellerMeier:1981}.  More recent
measurements of \nHO\ compared to \nHeO\
inside of the heliosphere find a similar ratio of H/He $\sim 6 - 7$
\citep{Richardsonetal:2004,GloecklerGeiss:2004,Witte_2004,Moebiusetal:2004}.
This difference can be attributed to two effects: the loss of 40--60\% of
interstellar \HO\ due to charge-transfer with protons in the heliosheath
region, a process denoted ``filtration'' \citep{RipkenFahr:1983}, and the
hardness of the interstellar radiation field at the Sun that ionizes more He
than H (\S \ref{sec:results}).

The \emph{Copernicus} satellite first showed that the local interstellar cloud
(LIC) surrounding the Sun is low density, $\sim 0.1$ atoms \cc, partially
ionized \citep[$n(\mathrm{H}^+)\sim n(\mathrm{H}^0)$,][]{York:1974,
McClintocketal:1975b}, and warm \citep[temperature $< 10^4$ K, e.g.][]
{McClintock_etal_1978}. \emph{Copernicus}, FUSE and HST data have shown that
the cluster of local interstellar clouds (CLIC, low density clouds within
$\sim 30$ pc), has low column densities, \NHI\ $< 10^{18.7}$, and N ionization
levels of $> 30$\% \citep[e.g.][]{Lehner_etal_2003,Wood_etal_2005}, indicating
partially ionized gas because H and N ionization are coupled by
charge-transfer.  The low column densities of the LIC itself, \NHI\ $< 10^{18}$
\cmtwo, indicate it is partially opaque to H ionizing radiation but not to He
ionizing radiation.  Cloud opacities of unity are reached for \NHI\ $\sim
10^{17.2}$ \cmtwo\ for photons close to the ionization threshold of hydrogen
(13.6 eV), and \NHI\ $\sim 10^{17.7}$ \cmtwo\ for photons at the \HeO\
ionization edge (24.6 eV). The result is that the LIC is partially ionized
with a significant ionization gradient between the edge and center.  Because
of this radiative transfer effects are important and need to be modeled
carefully in order to determine the boundary conditions of the heliosphere.

The LIC belongs to a flow of low density ISM embedded in the very low
density and apparently hot ($T \sim 10^6$ K) Local Bubble
\citep[][]{Frisch:1981,FrischYork:1986,McCammon_etal_1983,Snowden_etal_1990}.
The bulk motion of the CLIC through the local standard of
rest\footnote{Heliocentric motions are converted to the local standard of
rest, LSR, using the Standard solar apex motion.} corresponds to a velocity of
$-17.0$ to $-19.4$ \kms\ from the direction $\ell \sim 331^\circ$, $b \sim
-5^\circ$ \citep{FrischSlavin:2006book}.  This upwind direction is near the
center of the Loop I superbubble and the center of the ``ring'' shadow that has
been attributed to the merging of Loop I and the Local Bubble
\citep{FrischYork:1986,Frisch:2007,Egger:1995}.  Individual cloudlets with
distinct velocities are identified in this flow
\citep[][]{Lallementetal:1986,FGW:2002}. The velocity of the cloud feeding gas
into the heliosphere has been determined by the velocity of interstellar \HeO\
in the heliosphere measured by the GAS detector on Ulysses, $-26.3$ \kms\
\citep{Witte_2004}.

Curiously, absorption lines at the LIC velocity are not observed in the
nearest interstellar gas in the upwind hemisphere, such as towards the closest
star $\alpha$ Cen or towards 36 Oph located $\sim 5$ pc beyond the heliosphere
nose \citep{AdamsFrisch:1977,Landsmanetal:1984,LinskyWood:1996,Wood_etal_2000,
Wood36Oph:2000}.  This lack of an absorption component at the LIC velocity in
the closest stars in the upstream direction indicates that the Sun is near the
edge of the LIC, so the CHISM may vary over short distances (and hence
timescales). 

We are able to test for possible past variations by comparing the first
interplanetary \HO\ \lya\ glow spectrum obtained by \emph{Copernicus} in 1975
with Hubble Space Telescope observations of the \lya\ spectra obtained during
the mid-1990's solar minimum conditions.  The observed \HO\ velocity and
intensity has not varied to within uncertainties over the twenty-year period
separating these two sets of observations, so that the CHISM velocity field is
relatively smooth over spatial scales of $\sim 120$ AU in the downwind
direction \citep{FrischSlavin:2005}.  The 1975 \emph{Copernicus} data were
acquired in the direction corresponding to ecliptic longitudes of $\lambda =
264.3^\circ$, $\beta=+15.0^\circ$, or $\sim13.3^\circ$ from the most recent
upwind direction derived from SOHO \HI\ \lya\ data
\citep[][Q06]{QuemeraisetalA:2006,QuemeraisetalB:2006}.  The \emph{Copernicus}
look direction was just outside of the ``groove'' expected in the \lya\ glow
in the ecliptic during solar minimum.  The groove is caused by increased
charge-transfer in the solar wind current sheet, which has a small tilt during
minimum conditions \citep{Bzowski:2003}.  The velocity of the \lya\ profile
observed by \emph{Copernicus} corresponds to $-24.8 \pm 2.6$ \kms, after
correction to the SOHO upwind direction.  Since Q06 measured \HO\ upwind
velocities during the solar minimum years of 1996 and 1997 of $- 25.7 \pm 0.2$
\kms\ and $ -25.3 \pm 0.2$ \kms, which is consistent with the
\emph{Copernicus} results, we must arrive at the conclusion that the flow of
interstellar \HO\ into the heliosphere was relatively constant between the
years of 1975 and 1997, so that any observations of variations in the
interplanetary \lya\ glow properties must be due to solar activity properties
alone.  The thermal broadening of the \emph{Copernicus} spectrum corresponded
to a temperature of $\sim 5400$ K, however measurement uncertainties allowed
temperatures of up to $20,000$ K.

In this paper we present new photoionization models of the LIC and show they
reproduce both the densities of the ISM at the heliosphere and column
densities in the LIC component towards the star \object{$\epsilon$~CMa}.  In
earlier papers, photoionization models of both the ``LIC'' and ``Blue Cloud''
components observed towards \eCMa\ were used to constrain the models
\citep[SF02,][]{Frisch+Slavin_2003}.  SF02 grouped the properties of the LIC
and Blue Cloud, both of which are with 2.7 pc of the Sun because they are also
observed towards Sirius, $\sim 12^\circ$ from the $\epsilon$~CMa sightline.
The LIC velocity and density are sampled by observations of interstellar \HeI\
inside of the heliosphere \citep[$-26.3$ \kms,][]{Witte_2004}, and
\citet{Gry+Jenkins_2001} show that the properties of the LIC and Blue Cloud
differ somewhat.  The present study therefore focuses on obtaining the best
model of heliosphere boundary conditions by using only data on the LIC inside
of the heliosphere, and the LIC component towards $\epsilon$~CMa.  The present
study also benefits from new atomic rates for the critical
\MgII$\rightarrow$\MgI\ dielectronic recombination coefficient, improved
cooling rates in the radiative transfer code Cloudy, recent values for the
pickup ion densities at the termination shock, and recent values for solar
abundances.

The most truly unique quality of the LIC is that we are inside of it and
therefore have the ability to sample the cloud directly via \emph{in situ}
observations carried out by a variety of spacecraft within the Solar System.
Of all the available measurements of LIC gas flowing into the Solar System,
the observations of the density and temperature of neutral He are apparently
the most robust.  Helium, unlike hydrogen, undergoes little ionization or
heating in traversing the heliosheath regions, no deflection due to radiation
pressure, and is destroyed by photoionization and electron-impact ionization
within $\sim 1$ AU of the Sun, so we expect that the density and temperature
of He$^0$ derived from the observations in the Solar System are truly
representative of the values in the LIC \citep{Moebiusetal:2004}.  In this
paper we put a special emphasis on matching these \HeI\ data by determining
model parameters that yield close agreement with the $n(\mathrm{He}^0)$ and
$T(\mathrm{He}^0)$ data simultaneously.

\section{Photoionization Model Constraints and Assumptions
\label{sec:constraint}}

The primary data constraints on our photoionization models are the LIC
component column densities towards \eCMa, \citep[][hereafter
GJ01]{Gry+Jenkins_2001}, \emph{in situ} observations of \HeO\
\citep{Witte_2004,Moebiusetal:2004}, pickup ions
\citep[PUI,][]{GloecklerFisk:2007}, and anomalous cosmic rays
\citep[ACR,][]{CummingsStone:2002}.  The astronomical and \emph{in
situ} observational constraints are summarized in Table \ref{tab:obs}.

\begin{table}
\caption{Observational Constraints \label{tab:obs}}
\begin{center}
\begin{tabular}{llc}
\hline\hline
Observed &  Observed$^\mathrm{a}$ & Notes$^\mathrm{b}$ \\
Quantity &  Value &  \\ 
\hline
$N($\ion{C}{ii}$)$ (cm$^{-2}$) & $1.4 - 2.1\times10^{14}$ & 1, S \\
$N($\ion{C}{ii}$^*)$ (cm$^{-2}$) & $1.3 \pm 0.2 \times10^{12}$ & 1 \\
$N($\ion{C}{iv}$)$ (cm$^{-2}$) & $1.2\pm 0.3\times10^{12}$ & 1 \\
$N($\ion{N}{i}$)$ (cm$^{-2}$) & $1.70 \pm 0.05 \times10^{13}$ & 1 \\
$N($\ion{O}{i}$)$ (cm$^{-2}$) & $1.4^{+0.5}_{-0.2} \times10^{14}$ & 1, S  \\
$N($\ion{Mg}{i}$)$ (cm$^{-2}$) & $7 \pm 2 \times10^{9}$ & 1 \\
$N($\ion{Mg}{ii}$)$ (cm$^{-2}$) & $3.1 \pm 0.1 \times10^{12}$ & 1 \\
$N($\ion{Si}{ii}$)$ (cm$^{-2}$)  & $4.52\pm 0.2 \times10^{12}$ & 1 \\
$N($\ion{Si}{iii}$)$ (cm$^{-2}$) & $2.3\pm 0.2 \times10^{12}$ & 1 \\
$N($\ion{S}{ii}$)$ (cm$^{-2}$) & $8.6 \pm 2.1 \times10^{12}$ & 1 \\
$N($\ion{Fe}{ii}$)$ (cm$^{-2}$) & $1.35\pm 0.05 \times10^{12}$ & 1 \\
$N($\ion{H}{i}$)$/$N($\ion{He}{i}$)$ & $14\pm 0.4$$^\mathrm{c}$ & 2 \\
$T$ (K) & $6300 \pm 340$ & 4 \\
$n($He$^0)$ (cm$^{-3}$) & $0.015\pm 0.003$ & 4, $f = 1  ^\mathrm{d}$ \\
$n($N$^0)$$^\mathrm{e}$ (cm$^{-3}$) & $5.47 \pm 1.37 \times 10^{-6}$  &
3, $f = 0.68 - 0.95 ^\mathrm{d} $ \\
$n($O$^0)$$^\mathrm{e}$  (cm$^{-3}$) & $4.82 \pm 0.53 \times 10^{-5}$
& 3, $f = 0.64 - 0.99 ^\mathrm{d} $ \\
$n($Ne$^0)$$^\mathrm{e}$ (cm$^{-3}$) & $5.82 \pm 1.16 \times 10^{-6}$
& 3, $f = 0.84 - 0.95 ^\mathrm{d} $ \\
$n($Ar$^0)$$^\mathrm{e}$ (cm$^{-3}$)  & $1.63 \pm 0.73 \times 10^{-7}$
& 3, $f = 0.53 - 0.95 ^\mathrm{d} $  \\
\hline
\end{tabular}
\begin{list}{}{}
\item[$^\mathrm{a}$]$N($\ion{C}{ii}$^*)$, $N($\ion{N}{i}$)$,
$N($\ion{O}{i}$)$, $N($\ion{Mg}{ii}$)$, $N($\ion{Si}{ii}$)$,
$N($\ion{S}{ii}$)$ and $N($\ion{Fe}{ii}$)$ are used to constrain the input
abundances of the models.
\item[$^\mathrm{b}$]The first number indicates the reference (below), and
``S'' indicates that the column density is based on a saturated line.
Filtration factors ($f$, \S \ref{sec:insitu}) are listed for neutrals that
have been observed as pickup ions.  
\item[$^\mathrm{c}$]The uncertainty given is only that due to uncertainties
listed in \citet{Dupuis_etal_1995} for the observed \ion{H}{i} and \ion{He}{i}
column densities with the implicit assumption that the ratio is the same on
all lines of sight.  Given the substantial intrinsic variation in this ratio,
however, the quoted uncertainty must be regarded as a lower limit to the true
uncertainty.
\item[$^\mathrm{d}$]We assume in this paper that for He $f_\mathrm{He} = 1$,
and $f$ is not allowed to exceed 1.  
Heliosphere models predict the range $f_\mathrm{He} = 0.92 - 1$ for He 
and $f_\mathrm{H}=0.50-0.74$ for H
\citep[see filtration factors in ][]{CummingsStone:2002,Mueller:2004,
Mueller_etal_2007,Izmodenovetal:2004}.
\item[$^\mathrm{e}$]Note that the pickup ion ratios are for values at the
termination shock.  N$^0$, O$^0$, Ne$^0$, and Ar$^0$ densities need to be
corrected for filtration in the heliosheath regions (\sc \ref{sec:insitu}). 
\end{list}
\end{center}
References: (1) \citet{Gry+Jenkins_2001} Note that the values shown are
for the LIC component towards \eCMa.; (2) \citet{Dupuis_etal_1995}; (3)
\citet{GloecklerFisk:2007}; (4) \citet{Witte_2004}.
\end{table}


\subsection{Astronomical Constraints -- The LIC towards \eCMa\
\label{sec:constraintism} }

The data toward \eCMa\ of GJ01 show four separate velocity components detected
in several different ions including \CII, \SiIII, \SiII, and \MgII.  One of
these velocity components, with a heliocentric velocity of $\sim 17$ \kms, is
identified as the LIC.  Another is identified with a second local cloud, the
``Blue Cloud'' (BC) at $\sim 10$ \kms.  In our previous study of LIC ionization
(SF02) we noted that since the BC is close to the LIC, and both are detected
towards \object{Sirius} at 2.7 pc
\citep{Lallement_etal_1994,Hebrard_etal_1999}, perhaps actually abutting the
LIC, we should treat the two as a single cloud for this line of sight.  Counter
arguments to this line of reasoning include the fact that the BC apparently has
different properties than the LIC, though whether it is colder or hotter is
unclear \citep[see][GJ01]{Hebrard_etal_1999}.  For this reason and because we
wish to find the ionization models that best predict interstellar neutral
densities inside and outside of the heliosphere, we assume in this paper that
the LIC and BC are truly separate clouds and select only the LIC components
towards \eCMa\ to constrain the models.

In the context of these models, the best astronomical constraints on
the neutral ISM component are \NI, the saturated \OI\ line, and to
some extent \MgI\ (though the line is weak).  The electron density can
be deduced from the ratio \MgII/\MgI, which is determined by
photoionization and both dielectronic and radiative recombination, and
the excitation of the \CII\ fine-structure lines, \CII/\CIIstar.
However, the heavy saturation of the \CII\ 1335 \AA\ line in the ISM
limits the accuracy of the determination of $N$(\CII), so in this
paper we have de-emphasized \CII/\CIIstar\ as an ionization
diagnostic. For a discussion of the use of \CII/\CIIstar\ and other
observations in diagnosing the C abundance and C/S ratio see
\citet{Slavin+Frisch_2006}. Elements with first ionization potentials
(FIP) $ < 13.6$ eV (e.g.\ Mg, Si, S, and Fe) are generally almost
entirely singly ionized in the LIC and thus the column densities of
these ions are close to the total column densities for the elements.

\subsection{\emph{In situ} Constraints -- \HeO, Pickup Ions, and Anomalous
Cosmic Rays \label{sec:insitu}}

Neutral atoms in the LIC penetrate the outer heliosphere regions, and become
ionized primarily by charge-transfer with the solar wind ions, photoionization,
and electron impact ionization \citep{Rucinskietal:1996}.  The composition of
this neutral population reflects the partially ionized state of the LIC, rather
than indicating a pure FIP effect.  Thus the observed abundances of neutrals
representing high FIP elements, such as He, Ne, and Ar, as well as H, N, and O,
do not reflect their elemental abundances directly. As a result, these neutrals
provide an interesting and unique constraint on the photoionization models.
Once ionized, these interstellar wind particles form a population of ions with
a distinct velocity distribution that are ``picked up'' by the solar wind and
convected outwards, where they are measured by various spacecraft
\citep{Moebiusetal:2004,GloecklerGeiss:2004}.  These pickup ions (PUI)
are accelerated in the heliosheath region and form a population of cosmic rays
with an anomalous composition reflective of their origin as interstellar
neutrals in a partially ionized gas \citep{CummingsStone:2002}.  \emph{In situ}
observations of these byproducts of the ISM interaction with the heliosphere,
H, He, N, O, Ar, and Ne, provide a unique opportunity to constrain theoretical
models of an interstellar cloud using the combination of sightline-integrated
data, and data from a ``single'' spatial location, the heliosphere.

We adopt the Ulysses He measurements as the best set of constraints on
the ISM inside of the heliosphere.  The Ulysses satellite provides
direct measurements of interstellar \HeO\ at high ecliptic latitudes
throughout the solar cycle \citep[the GAS detector,][]{Witte_2004} and
also measurements of the He pickup ion component \citep[the SWICS
detector][]{GloecklerGeiss:2004,GloecklerGeiss:2006}.  Although
interstellar \HeO\ close to the Sun is detected through the resonant
scattering of solar 584 \AA\ radiation, geocoronal contamination of
the interstellar signal is present so we prefer the Ulysses data
\citep{Moebiusetal:2004}.  We adopt the Ulysses GAS and SWICS results,
\nHeO\ $ = 0.0151\pm0.0015$ cm$^{-3}$, $T(\mathrm{He}^0) = 6,300 \pm
340$ K. 

Data on the density of neutral N, O, Ne, and Ar in the surrounding ISM are
provided by PUI and ACR data.  The densities of interstellar N$^0$, O$^0$,
Ne$^0$, and Ar$^0$ at the termination shock are listed in Table \ref{tab:obs}.
These densities must be corrected by the filtration factors, which correspond
to the ratios of the densities at the termination shock to those in the LIC.
Filtration occurs when neutral interstellar atoms are removed from the inflow
by charge-transfer with interstellar protons as the atom crosses the
heliosheath regions.  Filtration values are listed in Table \ref{tab:obs},
based on values in \citet{CummingsStone:2002,MuellerZank:2004,
Izmodenovetal:2004}.  Filtration values larger than 1 are not considered,
although some models suggest possible net creation of O$^0$ through
charge-transfer between O$^+$ and H$^0$ in the heliosheath regions
\citep{Mueller:2004}.  We adopt $f_\mathrm{He} = 1$ for He.  Our models must
be compared to the interstellar densities obtained by correcting densities at
the termination shock by filtration factors.

\subsection{Interstellar Radiation Field at the Cloud Surface 
\label{sec:radfield}}

The spectrum and flux of the cosmic radiation field control the ionization of
the very local warm partially ionized medium (WPIM).  The local interstellar
radiation field (ISRF) that ionizes the LIC and other nearby cloudlets is
determined by the location of the Sun in the interior of a hole in the neutral
interstellar gas and dust referred to as the Local Bubble.  The clustering
tendency of hot O and B stars, and attenuation of radiation by interstellar
dust and gas, yield the well known spatial variation of the intensity and
spectrum of the ISRF.  We ignore possible temporal variations of the radiation
field \citep[e.g.][]{Parravano_etal_2003}, and model the ISRF throughout the
LIC based on the observations of the present-day radiation field.

\begin{figure}
\resizebox{\hsize}{!}{\includegraphics{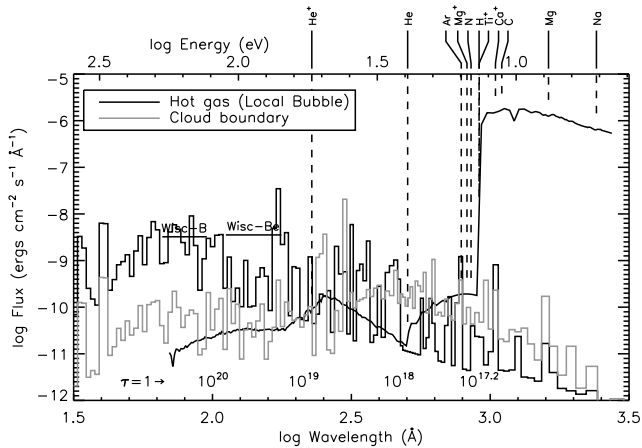}}
\caption{The modeled interstellar radiation field at the Sun (model 26) is
shown as a function of wavelength (bottom X-axis) and energy (top X-axis).
The black histogram is the modeled hot gas (i.e.\ Local Bubble) spectrum while
the gray histogram is the cloud boundary contribution.  The other line is the
stellar EUV/FUV background. The list of elements at the top of the plot
identifies the ionization potentials for neutrals and ions of interest.  The
energy/wavelength at which an optical depth of 1 is reached for several
different \ion{H}{i} column densities is shown along the bottom of the plot.
Observed flux levels of the soft X-ray diffuse background in the Wisconsin Be
and B bands are plotted as lines \citep{BlochSanders:1986,McCammon_etal_1983}.
}
\label{fig:radfield}
\end{figure}

The radiation field we use in our models is based on observations of the ISRF
at the Sun, supplemented by theoretical calculations of the spectra in the EUV and soft X-rays
where lack of sensitivity and/or spectral resolution require the use of models
to create a realistic spectrum.  The directly observed radiation field
includes the far ultraviolet (FUV) field created primarily by B stars, the
extreme ultraviolet (EUV) field from two B stars ($\epsilon$~CMa and
$\beta$~CMa) and hot nearby white-dwarf stars, and the diffuse soft X-ray
 background (SXRB).  We show the radiation field at the cloud surface in Fig.
\ref{fig:radfield}.  Because the interstellar opacity for radiation with $E >
13.6$ eV is vastly greater than for $E < 13.6$ eV, the EUV part of the ISRF
originates in nearby regions with $N($\ion{H}{i}$) \la 10^{18}$ \cmtwo\ while
the FUV comes from a much larger volume.

We use the EUV field of \citet{Vallerga_1998}, which is based on data
collected by the Extreme Ultraviolet Explorer (EUVE) satellite.  The EUVE
spectrometers were sensitive over the wavelength range of 504 -- 730 \AA\ and
showed that the stellar part of the EUV background is dominated by \eCMa\ and
$\beta$~CMa with substantial contributions from nearby hot white dwarfs at
shorter wavelengths. Vallerga extrapolated those measurements to the \HO\
ionization edge at 912 \AA\ using a total interstellar \HO\ density towards
\eCMa\ of \NHI\ $ = 9 \times 10^{17}$ \cmtwo.  This value for \NHI\ appears
somewhat high based on observations of \citet{Gry+Jenkins_2001} which, though
dependent on assumptions for gas phase abundances, indicate a value for \NHI\
of $\sim 7\times10^{17}$ \cmtwo.  This uncertainty in the total \NHI\ affects
only the extrapolated portion of the spectrum in our models since the other
portion of the spectrum is derived by de-absorbing the observed spectrum by
the value of \NHI\ assumed just for the LIC. For the results presented here we
have assumed that the total \NHI\ towards \eCMa\ is $7\times10^{17}$ \cmtwo.
Assuming a larger value would increase the flux in the extrapolated region,
though our calculations indicate that the overall affect on our results is
only at the $\sim 2$\% level at most.

The FUV field is important because it sets the ionization rate of Mg$^0$, which
has a first ionization potential of 7.65 eV (1621 \AA).  The radiation field
shortwards of 1600 \AA\ is heavily dominated by O and B stars in Gould's Belt,
particularly those occupying the unattenuated regions of the third and fourth
galactic quadrants.  A pronounced spatial asymmetry in the 1565 \AA\ radiation
field has been observed by the TD-1 satellite S2/68 telescope survey of the
interstellar radiation field, and we use those data and the extrapolation down
to 912 \AA\ from the 1564 \AA\ measurements as calculated by
\citet{Gondhalekar_etal_1980}.  The asymmetries in the TD-1 1565 \AA\
radiation field are reproduced by diffuse interstellar radiation field models
\citep{Henry_2002}.

The diffuse soft X-ray background (SXRB) has been observed over the entire sky
at relatively low spatial and spectral resolution by \emph{ROSAT}
\citep{Snowden_etal_1997} and proportional counters flown on sounding rockets
by the Wisconsin group \citep[e.g.,][]{McCammon_etal_1983}.  The broadband
count rates in the low energy bands, particularly the B and C bands (130--188
eV and 160--284 eV respectively) have been modeled as coming from an optically
thin, hot plasma at a temperature of $\sim 10^6$ K that occupies the low
density cavity extending to $\sim 50$--200 pc from the Sun in all directions
\citep{Snowden_etal_1990}.  Refinements to this picture have been required by
\emph{ROSAT} data showing absorption by relatively distant clouds \citep[e.g.\
MBM12,][]{Snowden_etal_1993}.  \citet{Snowden_etal_1998} propose a picture in
which the emission is divided between a LB component (unabsorbed except for
the LIC) and a distant absorbed component, mainly in the Galactic halo. More
recently there has been growing evidence that some, possibly large, fraction
of the SXRB is generated within the heliosphere from charge exchange between
solar wind ions and neutral atoms (SWCX).  We discuss this further in
\S\ref{sec:disc-rf}.

We model the spectrum measured by the broad-band soft X-ray observations by
assuming that the SXR background consists of a local (unabsorbed) component
and a distant (halo) component absorbed by an \HI\ column density of $10^{19}$
\cmtwo.  The emission measure or intrinsic intensity of the local and distant
components are assumed to be equal.  The spectrum is calculated using the
\citet[updated]{Raymond+Smith_1977} plasma emission code assuming a hot,
optically thin plasma in collisional ionization equilibrium. We explore
temperatures for the hot gas of $\log T_h = 5.9, 6.0$ and 6.1.  The total
flux, scaled by the emission measure, is fixed so that the B band flux
matches the all-sky average from \citet{McCammon_etal_1983}.

The boundary region between the warm LIC and the adjacent Local Bubble plasma
may be another significant source of EUV radiation and we include this flux in
our models 1--30.  We model this transition region as a conductive interface
between the LIC and Local Bubble plasma in the same way as described in
\citet{Slavin+Frisch_2002}.  The cloud is assumed to be steadily evaporating
into the surrounding hot gas.  The partially ionized gas of the LIC is heated
and ionized as it flows into the Local Bubble.  The ionization falls out of
equilibrium, with low ionization stages persisting into the hot gas.  The
non-equilibrium ionization is followed and the emission in the boundary is
calculated again using the \citet{Raymond+Smith_1977} code.  Ions in the
outflow are typically excited several times before being ionized and the
boundary region radiates strongly in the $13.6 - 54.4$ eV band.  The
contribution of the interface emission to the total B band flux is taken into
account in the calculation of the emission measure for the hot gas of the
Local Bubble so that the B band flux still matches the all-sky average.

We note that no attempt is made to make this model consistent with the size of
the local cavity, which is proposed to contain the hot gas in the standard
model for the SXRB \citep[e.g.,][]{Snowden_etal_1998}.  The pressure in the
hot gas is not adjusted to fit such models.  Rather the total pressure in the
hot gas for an evaporating cloud model is dictated by the assumed density,
temperature and magnetic field in the cloud.  In fact the pressures in our
models come out far too low for the standard model to produce the SXRB within
the confines of the Local Cavity as deduced, e.g., from \ion{Na}{i}
observations \citep{Lallement_etal_2003}.  This in turn means that if the
thermal pressure in the Local Bubble turns out to be much lower than was
assumed in those models because a substantial fraction of the SXRB comes from
SWCX then the emission from the cloud boundary would be unaffected.

\begin{figure}
\resizebox{\hsize}{!}{\includegraphics{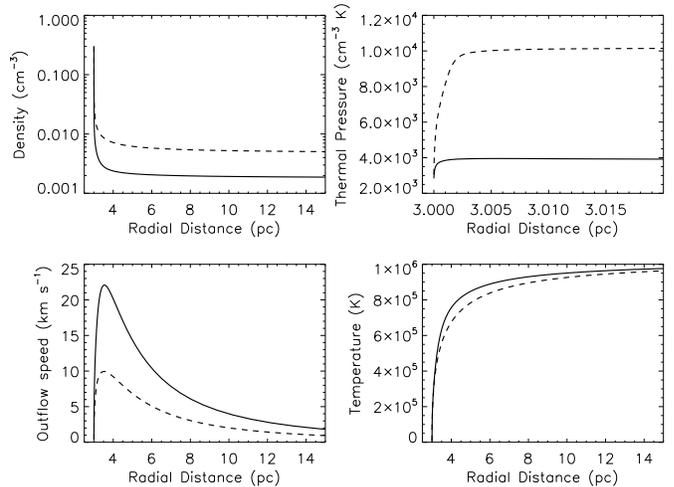}}
\caption{Profiles of hydrodynamic variables in two different cloud boundary
calculations.  The solid lines are for model 6 which has $n_\mathrm{H} =
0.273$ cm$^{-3}$, $\log T_h = 6.0$, $B_0 = 2\, \mu$G and $N_\mathrm{H I} = 4.5
\times 10^{17}$ cm$^{-2}$, while the dashed lines are for model 8, which
differs from model 6 only in the strength of the magnetic field, $B_0 = 5\,
\mu$G.  Note that in the upper right panel, the plot of thermal pressure, the
radial scale is much smaller in order to show the variation in thermal
pressure.  The higher magnetic pressure inside the cloud for model 8 leads to
the higher external thermal pressure for that case.  The temperature profile
differs in the two cases because the degree of heat flux saturation is reduced
for the higher thermal pressure of model 8, which in turn leads to a shallower
temperature gradient.}
\label{fig:cldprof}
\end{figure}

An example of the profiles of the hydrodynamic variables for two different
models is shown in Fig. \ref{fig:cldprof}.  The magnetic field strength in
these calculations is assumed to be proportional to the density at every point
in the outflow.  The treatment here is the same as in \citet{Slavin_1989}
which contains a more thorough discussion of the issues involved in this sort
of calculation.  The effect of the field on the conductivity is parametrized
in a simple way by a constant reduction factor of 0.5.  The importance of the
magnetic field for our calculations lies in the way it affects the thermal
pressure in the layer.  Since the density drops sharply in the outflow and
$|B| \sim n$, any magnetic pressure ($\sim B^2$) drops even more sharply.  The
total pressure is roughly constant in the boundary, so the thermal pressure
necessarily rises to make up for the decreasing magnetic pressure.  Since all
our models have nearly the same thermal pressure in the cloud, the primary
effect of the magnetic field is to help determine the thermal pressure in the
interface and thus radiative flux from the boundary.  Since the total soft
X-ray flux is fixed by requiring a match with the Wisconsin B band all-sky
average count rate \citep{McCammon_etal_1983}, a larger assumed magnetic field
increases only the EUV flux which is not constrained by the B-band data.  The
affect on the cloud of a larger EUV flux is to increase its temperature and
ionization.

A secondary effect of the magnetic field can be seen in the temperature
profiles in Figure \ref{fig:cldprof}.  In evaporating clouds in the ISM, if
the temperature gradient is large enough and the thermal pressure is low
enough, the conduction becomes ``saturated'', which means that the heat flux
expected from the gas temperature and temperature gradient exceeds the flux
that can be carried by the electrons \citep{Cowie+McKee_1977}.  Saturation
leads to a steepening of the temperature gradient and a relative slowing of
the mass loss rate.  In the two cases shown in Figure \ref{fig:cldprof}, the
lower magnetic field case ($B_0 = 2\,\mu$G) is moderately saturated (in terms
of Cowie \& McKee's parameter, $\sigma_0 = 3$) while the other case ($B_0 =
5\,\mu$G), because of the higher magnetic pressure, is less saturated.

\begin{table}
\caption{Model Input Parameter Values\label{tab:modparm}}
\centering
\begin{tabular}{cccccc} 
\hline\hline
 & \multicolumn{4}{c}{Input Parameter} \\ 
\cline{2-5} \\
 & $n_\mathrm{H}$ & $\log T_h$ & $B_0$ & $N_\mathrm{HI}$ \\ 
Model No. & (cm$^{-3}$) & (K) & ($\mu$G) & (10$^{17}$ cm$^{-2}$) \\
\hline
 1 & 0.273 &   5.9 &   2.0 &   3.0  \\
 2 & 0.273 &   5.9 &   2.0 &   4.5  \\
 3 & 0.273 &   5.9 &   5.0 &   3.0  \\
 4 & 0.273 &   5.9 &   5.0 &   4.5  \\
 5 & 0.273 &   6.0 &   2.0 &   3.0  \\
 6 & 0.273 &   6.0 &   2.0 &   4.5  \\
 7 & 0.273 &   6.0 &   5.0 &   3.0  \\
 8 & 0.273 &   6.0 &   5.0 &   4.5  \\
 9 & 0.273 &   6.1 &   2.0 &   3.0  \\
10 & 0.273 &   6.1 &   2.0 &   4.5  \\
11 & 0.273 &   6.1 &   5.0 &   3.0  \\
12 & 0.273 &   6.1 &   5.0 &   4.5  \\
13 & 0.218 &   5.9 &   2.0 &   3.0  \\
14 & 0.218 &   5.9 &   2.0 &   4.5  \\
15 & 0.218 &   5.9 &   5.0 &   3.0  \\
16 & 0.218 &   5.9 &   5.0 &   4.5  \\
17 & 0.218 &   6.0 &   2.0 &   3.0  \\
18 & 0.218 &   6.0 &   2.0 &   4.5  \\
19 & 0.218 &   6.0 &   5.0 &   3.0  \\
20 & 0.218 &   6.0 &   5.0 &   4.5  \\
21 & 0.218 &   6.1 &   2.0 &   3.0  \\
22 & 0.218 &   6.1 &   2.0 &   4.5  \\
23 & 0.218 &   6.1 &   5.0 &   3.0  \\
24 & 0.218 &   6.1 &   5.0 &   4.5  \\
25 & 0.226 &   5.9 &   4.7 &   3.0  \\
26 & 0.213 &   5.9 &   2.5 &   4.0  \\
27 & 0.226 &   6.0 &   3.8 &   3.0  \\
28 & 0.216 &   6.0 &   2.1 &   4.0  \\
29 & 0.232 &   6.1 &   3.4 &   3.0  \\
30 & 0.223 &   6.1 &  0.05 &   4.0  \\
42 & 0.218 &   6.1 &   --  &   4.5  \\
\hline
\end{tabular}
\end{table}

\section{Photoionization Models\label{sec:photmod}}

The photoionization models of the LIC are developed following the same
underlying procedure as in SF02, but selecting only the LIC absorption
component towards \eCMa\ for comparison.  Improvements include using updated
values for \emph{in situ} ISM observations and using a recent version of the
Cloudy radiative transfer/thermal equilibrium code \citep[version
06.02.09a,][]{Ferland_etal_1998}.  We run Cloudy with the assumption of a
plane-parallel cloud geometry and format our calculated photoionizing spectrum
to be used as input.  The selected options include utilizing recent
calculations for the dielectronic recombination rates for $\mathrm{Mg}^+
\rightarrow \mathrm{Mg}^0$ \citep[e.g.][and the commands ``set dielectronic
recombination Badnell'' and ``set radiative recombination
Badnell'']{Altun_etal_2006} , the assumption of constant pressure, and the
inclusion of interstellar dust grains at 50\% abundance compared to a standard
ISM value. As we discuss below, the only role that dust plays is net heating
of the gas since there is far too little column for extinction to be important
(expected $E(B-V) \sim 10 ^{-4.2}$).  The fraction of the heating provided by
dust is $\sim 4$\% of the total heating, while dust provides $\sim 2$\% of the
cooling (mainly by capture of electrons onto grain surfaces).  A cosmic ray
ionization rate at the default background level of $2.5\times10^{17}$ \psec\
(for H ionization) is included.  The LIC is assumed to be in ionization and
thermal equilibrium, and Cloudy calculates the detailed transfer of radiation,
including absorption and scattering of the radiation incident on the cloud
surface, as well as the diffuse continuum and emission lines generated within
the cloud.

The procedure for creating a model begins with generating the incident
radiation field at the cloud surface (\S \ref{sec:radfield}).  The
radiative transfer model is then run, and the output predictions of
the model are compared to observations of interstellar absorption
lines in the LIC towards \eCMa\ ($N($\ion{C}{ii}$^*)$,
$N($\ion{N}{i}$)$, $N($\ion{O}{i}$)$, $N($\ion{Mg}{ii}$)$,
$N($\ion{S}{ii}$)$, $N($\ion{Si}{ii}$)$, $N($\ion{Fe}{ii}$)$) and
\emph{in situ} observations of \nHeO\ by spacecraft inside of the
solar system.  The abundances of C, N, O, Mg, Si, S and Fe are then
adjusted to be consistent with the observed column densities.  With
these new abundances, the Cloudy run is repeated and this process is
continued until no adjustment of the abundances is needed.  Because
the abundances do have an impact on the emission from the cloud
boundary, the cloud evaporation model is then re-run with the new
abundances as well to re-generate the input ionizing spectrum.  The
iterative process of generating the spectrum and doing the Cloudy
photoionization runs generally requires only a couple runs of the
cloud evaporation program and a few runs of Cloudy.

\section{Model Results \label{sec:results}}

\begin{figure}
\resizebox{\hsize}{!}{\includegraphics{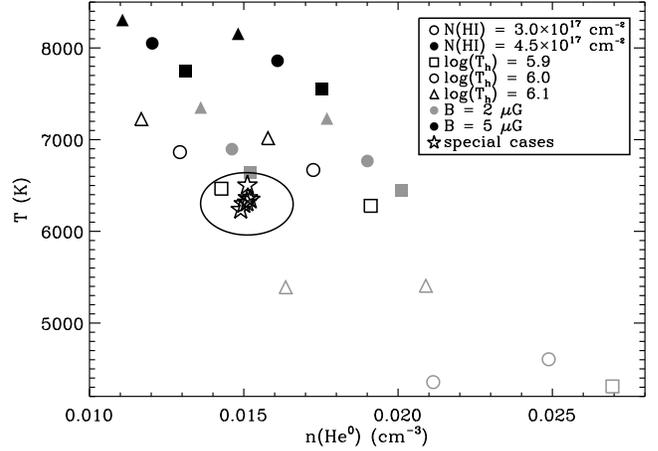}}
\caption{Model results for the He$^0$ density and temperature in the ISM just
outside the heliosphere.  The squares, circles, and triangles are for models
that are part of the initial grid of 24 models, while the stars are for models
$25-30$ for which the magnetic field, $B$, and the total H density,
$n_\mathrm{H}$, were varied to match the observed \nHeO\ and $T$. For the grid
models, the empty symbols are for models with \NHI$ = 3\times10^{17}$ \cmtwo,
while the filled symbol models have \NHI$ = 4.5\times10^{17}$ \cmtwo. As the
legend shows, the color (black vs.\ gray) indicates the magnetic field
strength and the symbol shape indicates the temperature assumed for the hot
gas of the Local Bubble.  For identical kinds of points, the one to the left
is for a model with $n(\mathrm{H}^0) = 0.218$ \cc\ and the one to the right
has $n(\mathrm{H}^0) = 0.273$ \cc.  The ellipse is the error range around the
observed values for $n(\mathrm{He}^0)$ and $T$.}
\label{fig:nHeT}
\end{figure}

\begin{table*}
\caption{Model Column Density Results\label{tab:coldens}} 
\centering
\begin{tabular}{cccccccc}
\hline\hline
Model$^\mathrm{a}$ & $\log N(\mathrm{H_{tot}})$ & $\log N($\ion{Ar}{i}$)$ & 
$\log N($\ion{Ar}{ii}$)$ & $\log N($\ion{Si}{iii}$)$ &
$\frac{N(\mathrm{Mg}\;\mathrm{II})}{N(\mathrm{Mg}\;\mathrm{I})}$ &
$\frac{N(\mathrm{C}\;\mathrm{II})}{N(\mathrm{C}\;\mathrm{II}^*)}$ &
$\frac{N(\mathrm{H}\;\mathrm{I})}{N(\mathrm{He}\;\mathrm{I})}$ \\
\hline
Obs.$^\mathrm{b}$ & -- & -- & -- & 12.40 & $443^{+197}_{-110}$ & $93 -
430^\mathrm{c}$ & $12 - 16$\\
\hline
1 & 17.58 & 11.48 & 11.76 & 8.895 & 706.3 & 213.0 & 11.36 \\
2 & 17.78 & 11.67 & 11.95 & 9.641 & 325.3 & 234.1 & 11.40 \\
3 & 17.65 & 11.33 & 11.84 & 9.844 & 246.1 & 182.1 & 11.44 \\
4 & 17.82 & 11.55 & 12.00 & 10.12 & 136.5 & 198.0 & 11.80 \\
5 & 17.58 & 11.49 & 11.75 & 8.961 & 717.8 & 230.0 & 11.62 \\
6 & 17.78 & 11.67 & 11.94 & 9.711 & 286.7 & 244.9 & 11.62 \\
7 & 17.64 & 11.34 & 11.81 & 9.909 & 207.4 & 194.8 & 12.47 \\
8 & 17.82 & 11.56 & 11.97 & 10.17 & 119.7 & 208.2 & 12.66 \\
9 & 17.59 & 11.46 & 11.75 & 9.289 & 553.0 & 239.5 & 12.27 \\
10 & 17.78 & 11.66 & 11.93 & 9.861 & 210.3 & 248.1 & 12.14 \\
11 & 17.64 & 11.33 & 11.78 & 9.993 & 171.2 & 201.7 & 13.44 \\
12 & 17.82 & 11.54 & 11.95 & 10.23 & 104.3 & 210.6 & 13.52 \\
13 & 17.60 & 11.44 & 11.78 & 9.100 & 768.9 & 255.1 & 11.52 \\
\textbf{14} & 17.79 & 11.64 & 11.96 & 9.738 & 336.4 & 274.4 & 11.47 \\
15 & 17.67 & 11.30 & 11.85 & 9.944 & 251.2 & 215.9 & 11.65 \\
16 & 17.84 & 11.51 & 12.01 & 10.21 & 140.6 & 232.1 & 12.01 \\
17 & 17.58 & 11.47 & 11.76 & 8.926 & 840.1 & 262.5 & 11.65 \\
18 & 17.79 & 11.65 & 11.95 & 9.792 & 308.9 & 286.3 & 11.64 \\
19 & 17.66 & 11.30 & 11.82 & 10.02 & 208.5 & 229.2 & 12.76 \\
20 & 17.84 & 11.52 & 11.98 & 10.26 & 123.5 & 242.1 & 12.95 \\
21 & 17.59 & 11.43 & 11.76 & 9.318 & 633.1 & 280.8 & 12.41 \\
22 & 17.79 & 11.63 & 11.94 & 9.926 & 224.9 & 288.4 & 12.25 \\
23 & 17.66 & 11.29 & 11.79 & 10.10 & 175.0 & 235.5 & 13.81 \\
24 & 17.84 & 11.50 & 11.96 & 10.31 & 110.2 & 245.5 & 13.85 \\
25 & 17.66 & 11.32 & 11.84 & 9.887 & 271.3 & 213.6 & 11.71 \\
\textbf{26} & 17.74 & 11.56 & 11.91 & 9.676 & 383.9 & 270.1 & 11.70 \\
\textbf{27} & 17.63 & 11.36 & 11.79 & 9.751 & 331.1 & 242.0 & 12.51 \\
\textbf{28} & 17.73 & 11.59 & 11.90 & 9.625 & 411.6 & 284.5 & 11.79 \\
\textbf{29} & 17.62 & 11.38 & 11.77 & 9.720 & 329.3 & 251.6 & 13.01 \\
\textbf{30} & 17.72 & 11.60 & 11.88 & 9.642 & 386.6 & 295.1 & 12.08 \\
\textbf{42} & 17.77 & 11.70 & 11.92 & 9.555 & 482.9 & 317.9 & 11.51 \\
\hline
\end{tabular}
\begin{list}{}{}
\item[$^\mathrm{a}$]The best models, consistent with all observational data
(see \S\ref{sec:results}), are indicated by bold face.
\item[$^\mathrm{b}$]Observational results from \citet{Gry+Jenkins_2001} (see
table \ref{tab:obs}). The values listed for $N($\ion{H}{i}$)/N($\ion{He}{i}$)$
are the range of values observed excluding Feige 24 which is one of the most
distant stars observed by \citet{Dupuis_etal_1995} and has unusually large
$N($\ion{H}{i}$)$ and ratio values.
\item[$^\mathrm{c}$]The upper limit on $N($\ion{C}{ii}$)$ is not well
determined observationally because of the saturation of the line.
\citet{Gry+Jenkins_2001} define it by assuming a solar C/S abundance ratio and
using the observed \ion{S}{ii} column density.  We find
\citep[see][]{Slavin+Frisch_2006} that the abundance of C required to match
the $N($\ion{C}{ii}$^*)$ observations is supersolar, with C/S$ \sim 36 - 44$,
which results in a much higher upper limit on $N($\ion{C}{ii}$)$.
\end{list}
\end{table*}

\begin{table*}
\caption{Model Results for Solar Location\label{tab:sunres}}
\centering
\begin{tabular}{cccccccccccc}
\hline\hline
Model$^\mathrm{a}$ & $X(\mathrm{H})$ & $X(\mathrm{He})$ & $n($H$^0)$ &
$n($He$^0)$ & $n($N$^0)$ & $n($O$^0)$ &
$n($Ne$^0)$ & $n($Ar$^0)$ & $n_e$ & $n_p$ & $T$ \\
\hline
Obs.$^\mathrm{b}$ & & & & 0.015 & $5.5\times10^{-6}$ & $4.8\times10^{-5}$ & 
$5.8\times10^{-6}$ & $1.6\times10^{-7}$ & & & 6300 \\
\hline
1 & 0.176 & 0.299 & 0.318 & 0.0269 & $1.84\times10^{-5}$ & $1.48\times10^{-4}$
& $1.39\times10^{-5}$ & $3.47\times10^{-7}$ & 0.0800 & 0.0678 & 4310 \\
2 & 0.196 & 0.352 & 0.251 & 0.0201 & $9.63\times10^{-6}$ & $7.71\times10^{-5}$
& $8.51\times10^{-6}$ & $2.69\times10^{-7}$ & 0.0724 & 0.0611 & 6450 \\
3 & 0.286 & 0.404 & 0.230 & 0.0191 & $1.34\times10^{-5}$ & $1.07\times10^{-4}$
& $7.81\times10^{-6}$ & $1.82\times10^{-7}$ & 0.106 & 0.0922 & 6280 \\
4 & 0.259 & 0.428 & 0.229 & 0.0175 & $8.86\times10^{-6}$ & $6.95\times10^{-5}$
& $6.45\times10^{-6}$ & $1.92\times10^{-7}$ & 0.0937 & 0.0799 & 7560 \\
5 & 0.174 & 0.310 & 0.300 & 0.0249 & $1.72\times10^{-5}$ & $1.39\times10^{-4}$
& $1.13\times10^{-5}$ & $3.30\times10^{-7}$ & 0.0750 & 0.0631 & 4610 \\
6 & 0.197 & 0.363 & 0.242 & 0.0190 & $9.13\times10^{-6}$ & $7.44\times10^{-5}$
& $7.08\times10^{-6}$ & $2.61\times10^{-7}$ & 0.0707 & 0.0593 & 6770 \\
7 & 0.278 & 0.441 & 0.225 & 0.0173 & $1.33\times10^{-5}$ & $1.04\times10^{-4}$
& $5.75\times10^{-6}$ & $1.78\times10^{-7}$ & 0.101 & 0.0866 & 6670 \\
8 & 0.257 & 0.459 & 0.223 & 0.0161 & $8.65\times10^{-6}$ & $6.79\times10^{-5}$
& $4.92\times10^{-6}$ & $1.89\times10^{-7}$ & 0.0918 & 0.0773 & 7860 \\
9 & 0.193 & 0.357 & 0.265 & 0.0209 & $1.53\times10^{-5}$ & $1.23\times10^{-4}$
& $7.52\times10^{-6}$ & $2.71\times10^{-7}$ & 0.0761 & 0.0634 & 5410 \\
10 & 0.203 & 0.392 & 0.235 & 0.0177 & $9.06\times10^{-6}$ &
$7.24\times10^{-5}$ & $5.52\times10^{-6}$ & $2.43\times10^{-7}$ & 0.0723 &
0.0600 & 7230 \\
11 & 0.278 & 0.474 & 0.220 & 0.0158 & $1.29\times10^{-5}$ &
$1.02\times10^{-4}$ & $4.42\times10^{-6}$ & $1.69\times10^{-7}$ & 0.0999 &
0.0847 & 7020 \\
12 & 0.262 & 0.490 & 0.219 & 0.0148 & $8.41\times10^{-6}$ &
$6.66\times10^{-5}$ & $3.83\times10^{-6}$ & $1.78\times10^{-7}$ & 0.0928 &
0.0776 & 8160 \\
13 & 0.201 & 0.333 & 0.230 & 0.0191 & $1.34\times10^{-5}$ &
$1.08\times10^{-4}$ & $8.94\times10^{-6}$ & $2.32\times10^{-7}$ & 0.0681 &
0.0580 & 4720 \\
\textbf{14} & 0.215 & 0.376 & 0.193 & 0.0152 & $7.29\times10^{-6}$ &
$5.95\times10^{-5}$ & $5.99\times10^{-6}$ & $1.94\times10^{-7}$ & 0.0625 &
0.0528 & 6650 \\
15 & 0.310 & 0.436 & 0.176 & 0.0143 & $1.02\times10^{-5}$ &
$8.18\times10^{-5}$ & $5.35\times10^{-6}$ & $1.28\times10^{-7}$ & 0.0905 &
0.0789 & 6470 \\
16 & 0.284 & 0.461 & 0.175 & 0.0131 & $6.76\times10^{-6}$ &
$5.36\times10^{-5}$ & $4.40\times10^{-6}$ & $1.35\times10^{-7}$ & 0.0812 &
0.0695 & 7750 \\
17 & 0.178 & 0.315 & 0.255 & 0.0211 & $1.47\times10^{-5}$ &
$1.19\times10^{-4}$ & $9.51\times10^{-6}$ & $2.75\times10^{-7}$ & 0.0656 &
0.0553 & 4360 \\
18 & 0.214 & 0.383 & 0.188 & 0.0146 & $7.16\times10^{-6}$ &
$5.81\times10^{-5}$ & $5.07\times10^{-6}$ & $1.91\times10^{-7}$ & 0.0610 &
0.0513 & 6900 \\
19 & 0.302 & 0.472 & 0.173 & 0.0129 & $1.01\times10^{-5}$ &
$8.03\times10^{-5}$ & $3.93\times10^{-6}$ & $1.26\times10^{-7}$ & 0.0872 &
0.0748 & 6860 \\
20 & 0.282 & 0.491 & 0.172 & 0.0120 & $6.61\times10^{-6}$ &
$5.26\times10^{-5}$ & $3.37\times10^{-6}$ & $1.33\times10^{-7}$ & 0.0802 &
0.0677 & 8050 \\
21 & 0.204 & 0.374 & 0.211 & 0.0164 & $1.20\times10^{-5}$ &
$9.85\times10^{-5}$ & $5.54\times10^{-6}$ & $2.04\times10^{-7}$ & 0.0648 &
0.0541 & 5390 \\
22 & 0.221 & 0.414 & 0.184 & 0.0136 & $6.96\times10^{-6}$ &
$5.69\times10^{-5}$ & $3.90\times10^{-6}$ & $1.78\times10^{-7}$ & 0.0628 &
0.0523 & 7350 \\
23 & 0.305 & 0.506 & 0.168 & 0.0117 & $9.91\times10^{-6}$ &
$7.82\times10^{-5}$ & $2.97\times10^{-6}$ & $1.18\times10^{-7}$ & 0.0868 &
0.0737 & 7220 \\
24 & 0.286 & 0.520 & 0.169 & 0.0111 & $6.57\times10^{-6}$ &
$5.16\times10^{-5}$ & $2.60\times10^{-6}$ & $1.25\times10^{-7}$ & 0.0808 &
0.0676 & 8310 \\
25 & 0.297 & 0.427 & 0.187 & 0.0151 & $1.09\times10^{-5}$ &
$8.67\times10^{-5}$ & $5.75\times10^{-6}$ & $1.41\times10^{-7}$ & 0.0907 &
0.0789 & 6360 \\
\textbf{26} & 0.224 & 0.385 & 0.192 & 0.0151 & $8.32\times10^{-6}$ &
$6.66\times10^{-5}$ & $5.96\times10^{-6}$ & $1.83\times10^{-7}$ & 0.0653 &
0.0554 & 6320 \\
\textbf{27} & 0.258 & 0.426 & 0.195 & 0.0149 & $1.13\times10^{-5}$ &
$8.97\times10^{-5}$ & $5.03\times10^{-6}$ & $1.63\times10^{-7}$ & 0.0796 &
0.0677 & 6240 \\
\textbf{28} & 0.212 & 0.376 & 0.194 & 0.0152 & $8.24\times10^{-6}$ &
$6.73\times10^{-5}$ & $5.49\times10^{-6}$ & $1.95\times10^{-7}$ & 0.0622 &
0.0523 & 6350 \\
\textbf{29} & 0.242 & 0.429 & 0.202 & 0.0150 & $1.17\times10^{-5}$ &
$9.30\times10^{-5}$ & $4.53\times10^{-6}$ & $1.74\times10^{-7}$ & 0.0769 &
0.0646 & 6300 \\
\textbf{30} & 0.203 & 0.379 & 0.197 & 0.0151 & $8.51\times10^{-6}$ &
$6.81\times10^{-5}$ & $4.78\times10^{-6}$ & $2.01\times10^{-7}$ & 0.0605 &
0.0502 & 6500 \\
\textbf{42} & 0.189 & 0.355 & 0.186 & 0.0145 & $6.91\times10^{-6}$ & 
$5.71\times10^{-5}$ & $4.60\times10^{-6}$ & $2.04\times10^{-7}$ & 0.0522 &
0.0433 & 6590 \\ 
\hline
\end{tabular}
\begin{list}{}{}
\item[$^\mathrm{a}$]The best models (see \S\ref{sec:results}) are indicated by
bold face.  \item[$^\mathrm{b}$]Observational results from
\citet{GloecklerGeiss:2004},
Gloeckler (2005, private communication) and \citet{Witte_etal_1996} (see table
\ref{tab:obs} for uncertainties).
\end{list}
\end{table*}

To investigate the dependence of the results on the input parameters
we calculate a grid of 24 models.  We explore total H density,
$n_\mathrm{H} = 0.273, 0.218$; Local Bubble hot gas temperature, $\log
T_h = 5.9, 6.0, 6.1$; magnetic field strength, $B_0 = 2, 5\,\mu$G; and
\ion{H}{i} column density, $N_\mathrm{H\,I} = 3\times10^{17},
4.5\times10^{17}$ \cmtwo.  The values for $T$ and $n($\ion{He}{i}$)$
at the solar location, the endpoint of the calculations, are shown in
Fig. \ref{fig:nHeT} for this set of models.  We then explore
another six models in which we do the calculations over a grid in
$\log T_h$ and $N_\mathrm{H\,I}$ but vary the values of $n_\mathrm{H}$
and $B_0$ in order to match the observed $T$ and $n($\ion{He}{i}$)$.
We employed a multiple linear regression to assist in narrowing down
the search region for the values of $n_\mathrm{H}$ and $B_0$ needed to
match the observations. Since the dependencies of the results on the
parameters really are quite non-linear, this procedure could not work
to predict exactly correct values for the required parameters, but was
useful for getting close to the correct values.  Based on the results
for the initial grid of models, we use $\log T_h = 5.9, 6.0, 6.1$ and
$N_\mathrm{H\,I} = 3\times10^{17}, 4\times10^{17}$ \cmtwo\ for this
smaller grid, models $25-30$.  We chose to use $4\times10^{17}$
\cmtwo\ rather than $4.5\times10^{17}$ \cmtwo\ because the higher
column density models, for the most part, produced temperatures that
were too high. As can be seen from Fig.\ \ref{fig:nHeT}, all of these
models (25--30), plotted as stars, are consistent with the observed
$T$ and $n($\ion{He}{i}$)$, indicated by the ellipse in the figure.
Table \ref{tab:modparm} gives the input parameters for each model.
Model predictions for the \eCMa\ sightline integrated through the LIC
are presented in Table \ref{tab:coldens}.  Model predictions for the
CHISM (i.e.\ at the solar location) are shown in Table \ref{tab:sunres}.

In SF02 we tested models with no emission from the cloud
boundary. This amounts to assuming that the boundary is a sharp
transition from the hot gas of the Local Bubble to the warm gas of the
LIC.  We have again explored such models using the LIC data as
constraints in the same way as for the models discussed above with a
conductive interface.  When there is no evaporative boundary, our
models do not depend on the magnetic field in the cloud since in that
case the ionizing flux consists only of diffuse emission from the hot
gas of the Local Bubble and EUV emission from nearby hot stars and
the spectrum is not related to the properties of the cloud.  For these
models our model grid consists of total H density, $n_\mathrm{H} =
0.273, 0.218$, Local Bubble hot gas temperature, $\log T_h = 5.9, 6.0,
6.1$, and \ion{H}{i} column density, $N_\mathrm{H\,I} =
3\times10^{17}, 4.5\times10^{17}$ \cmtwo, for a total of twelve models
that we label as models 31--42.  Of these models, however, only two
resulted in ionizing fluxes sufficient to heat the cloud to a
temperature $\sim 6000$ K at the same time as matching the constraints
on the ion column densities.  These models were the ones with $\log
T_h = 6.1$ and $N_\mathrm{H\,I} = 4.5\times10^{17}$ \cmtwo\ (models 36
and 42).  For the other cases, either at the surface or deeper into
the cloud, there are insufficient photons to provide the heating to
balance the cooling and the cloud temperature drops sharply to $< 1000$
K.  The model with $n_\mathrm{H} = 0.218$ (model 42) is consistent
with the observed value of \nHeO\ and with the column density data as
well.

From these 42 models we have selected those that provide acceptable
results for the observational constraints, according to prioritized
requirements.  The first requirement is that the model predict the
density and temperature of \HeI\ observed inside of the solar system.
Models 14 (marginally), 15 and 25--30 and 42 predict a He density and
temperature consistent with the observed values within the reported
errors.  They also predict the PUI Ne densities, for an assumed
Ne/H=123 ppm.  Models 26 and 28 successfully match the PUI data for
Ne/H as low as $\sim 100$ ppm.  These models are also required to
match the observed \MgII/\MgI\ ratio in the LIC towards \eCMa.  Models
14, 27, and 29 marginally fit this criterion, while 26, 28, 30, and 42
successfully fit this criterion.  Note that the new models now provide
acceptable predictions for the CHISM temperature, which was not the case
for the best models in SF02.  These models are also consistent with
the observed \CII/\CIIstar\ ratios in the LIC component towards \eCMa\
though this constraint is weak because of the large uncertainties in
$N$(\CII).  Based on these comparisons, and given the uncertainties in
both data and models, Models 14, 26-30, and 42 are plausible models,
but Models 26 and 28 appear to best match the observational
constraints.

Based on the constraints and assumptions presented in the previous
section, we select models 14, 26--30 and 42 as the best models for the
LIC ionization, with models 26 and 28 favored by the PUI Ne data
provided that Ne/H $> 100$ ppm.  We believe that these seven models
then give a realistic range for the uncertainties in the boundary
conditions of the heliosphere, providing that the underlying
assumptions implicit in the Cloudy code, e.g.\ photoionization
equilibrium, are correct.  The predictions of the best models provide
excellent matches to the observational constraints.  Based on these
models, we find the boundary conditions of the heliosphere to be
describable as \nHO\ $ = 0.19 - 0.20$ \cc, \nel\ $ = 0.05 - 0.08$ \cc,
and $X(\mathrm{H}) \equiv \mathrm{H}^+/(\mathrm{H}^0 +
\mathrm{H}^+) = 0.19 - 0.26$, $X(\mathrm{He}) \equiv
\mathrm{He}^+/(\mathrm{He}^0 + \mathrm{He}^+) = 0.36 - 0.43$.  For
these models, we find abundances of O/H $= 295-437$ ppm, C/H $= 589 -
813$ ppm, and N/H $= 40.7 - 64.6$ ppm (Table \ref{tab:abund}).  The
total LIC density is $n(\mathrm{H})_0 = 0.213 - 0.232$ \cc, while the
strength of the interstellar magnetic in the cloud varies between 0
and 3.8 \mG.  The Ne PUI data further favor densities of \nHO\ $\approx 0.19 $
\cc\ and \nel\ $ = 0.06 - 0.07$ \cc.

In this analysis we have assumed negligible filtration for He in the
heliosheath regions.  Modeling of the He filtration factor however allows
values as small as $f_\mathrm{He} = 0.92$, which yields \nHeI =  0.0164 \cc\
for the CHISM (Table \ref{tab:obs}).  The predictions of Model 21 agree with
this value for \nHeI, as well as with the ratios \MgII/\MgI\ and
\CII/\CIIstar\ towards $\epsilon$ CMa and the pickup ion Ne and Ar data (Table
\ref{tab:coldens}).  The predicted cloud temperature is low by $\sim 1000$ K.
The density and ionization for Model 21 are \nHO\ $= 0.21$ \cc\ and
\nel $=0.06$ \cc.  We therefore conclude that our best models listed above are
robust in the sense that they predict consistent values for the
\nHO\ to within 5\%, and electron densities to within 25\%.

The radiation field incident on the LIC for Model 26 is shown in Fig.
\ref{fig:radfield} and the spectral characteristics of the field for each
model are listed in Table \ref{tab:spect_charac}.  The wavelengths regions
$\lambda \le 912$ \AA\ and $\lambda \sim 1500 $ \AA\ are of primary
importance for the photoionization of the cloud, the former because it
determines \HO\ and \HeO\ ionization, and the latter because it determines
Mg$^0$ ionization.

The ionization parameter is defined as $U \equiv \Phi/n(\mathrm{H}) c $, where
$\Phi$ is the H ionizing photon flux, $n(\mathrm{H})$ is the total (neutral +
ionized) hydrogen density at the cloud surface and $c$ is the speed of light.
The total ionizing photon fluxes at the cloud surface (photons \cmtwo\
s$^{-1}$) for the three bands 13.6--24.6, 24.6--54.4, and 54.4--100 eV, are
given by $\Phi_\mathrm{H}$, $\Phi_\mathrm{He^0}$, and $\Phi_\mathrm{He^+}$,
respectively.  The ratio of the total number of \HO\ and \HeO\ ionizing
photons in the incident radiation field is given by $Q$(\HeO)/$Q$(\HO). The
quantity $\langle E\rangle$ (eV) is the mean energy of an ionizing photon,
equal to the integrated energy flux from 13.6 to 100 eV divided by the
integrated photon flux over the same energy range.

\begin{table*}
\caption{Characteristics of the Model Radiation Field\label{tab:spect_charac}}
\centering
\begin{tabular}{ccccccc}
\hline\hline
Model$^\mathrm{a}$ & U & $\phi_\mathrm{H}$ & $\phi_{\mathrm{He}^0}$ &
$\phi_{\mathrm{He}^+}$ & $Q$(He$^0$)/$Q$(H$^0$) & $\langle E \rangle$ \\
      &   & photons \cmtwo\ s$^{-1}$  & photons \cmtwo\ s$^{-1}$  &
photons \cmtwo\ s$^{-1}$  &  &  eV  \\
\hline
1 & $2.0\times10^{-6}$ & $4.6\times10^{3}$ & $7.5\times10^{3}$ &
$2.8\times10^{3}$ & 0.46 & 74.6 \\
2 & $2.0\times10^{-6}$ & $5.3\times10^{3}$ & $7.3\times10^{3}$ &
$2.7\times10^{3}$ & 0.44 & 72.4 \\
3 & $3.1\times10^{-6}$ & $9.0\times10^{3}$ & $1.3\times10^{4}$ &
$2.7\times10^{3}$ & 0.49 & 57.7 \\
4 & $3.0\times10^{-6}$ & $8.6\times10^{3}$ & $1.2\times10^{4}$ &
$2.6\times10^{3}$ & 0.49 & 58.6 \\
5 & $2.0\times10^{-6}$ & $4.0\times10^{3}$ & $7.0\times10^{3}$ &
$3.8\times10^{3}$ & 0.43 & 79.0 \\
6 & $2.1\times10^{-6}$ & $4.9\times10^{3}$ & $6.9\times10^{3}$ &
$3.8\times10^{3}$ & 0.41 & 76.2 \\
7 & $3.2\times10^{-6}$ & $7.4\times10^{3}$ & $1.4\times10^{4}$ &
$3.8\times10^{3}$ & 0.53 & 61.3 \\
8 & $3.1\times10^{-6}$ & $7.4\times10^{3}$ & $1.3\times10^{4}$ &
$3.7\times10^{3}$ & 0.52 & 61.5 \\
9 & $2.2\times10^{-6}$ & $3.8\times10^{3}$ & $7.4\times10^{3}$ &
$5.6\times10^{3}$ & 0.40 & 80.0 \\
10 & $2.3\times10^{-6}$ & $4.8\times10^{3}$ & $7.2\times10^{3}$ &
$5.6\times10^{3}$ & 0.38 & 77.3 \\
11 & $3.5\times10^{-6}$ & $6.6\times10^{3}$ & $1.5\times10^{4}$ &
$5.6\times10^{3}$ & 0.53 & 63.4 \\
12 & $3.4\times10^{-6}$ & $6.9\times10^{3}$ & $1.5\times10^{4}$ &
$5.5\times10^{3}$ & 0.52 & 63.2 \\
13 & $2.3\times10^{-6}$ & $4.3\times10^{3}$ & $7.0\times10^{3}$ &
$2.8\times10^{3}$ & 0.46 & 76.9 \\
\textbf{14} & $2.4\times10^{-6}$ & $5.1\times10^{3}$ & $6.8\times10^{3}$ &
$2.7\times10^{3}$ & 0.43 & 74.3 \\
15 & $3.7\times10^{-6}$ & $8.5\times10^{3}$ & $1.2\times10^{4}$ &
$2.7\times10^{3}$ & 0.49 & 58.8 \\
16 & $3.6\times10^{-6}$ & $8.2\times10^{3}$ & $1.2\times10^{4}$ &
$2.6\times10^{3}$ & 0.49 & 59.5 \\
17 & $2.3\times10^{-6}$ & $3.7\times10^{3}$ & $6.4\times10^{3}$ &
$3.8\times10^{3}$ & 0.42 & 81.8 \\
18 & $2.4\times10^{-6}$ & $4.7\times10^{3}$ & $6.2\times10^{3}$ &
$3.8\times10^{3}$ & 0.39 & 78.4 \\
19 & $3.9\times10^{-6}$ & $7.1\times10^{3}$ & $1.4\times10^{4}$ &
$3.8\times10^{3}$ & 0.53 & 62.2 \\
20 & $3.8\times10^{-6}$ & $7.2\times10^{3}$ & $1.3\times10^{4}$ &
$3.7\times10^{3}$ & 0.52 & 62.2 \\
21 & $2.6\times10^{-6}$ & $3.6\times10^{3}$ & $6.8\times10^{3}$ &
$5.6\times10^{3}$ & 0.39 & 82.0 \\
22 & $2.7\times10^{-6}$ & $4.6\times10^{3}$ & $6.6\times10^{3}$ &
$5.6\times10^{3}$ & 0.36 & 78.9 \\
23 & $4.2\times10^{-6}$ & $6.3\times10^{3}$ & $1.5\times10^{4}$ &
$5.6\times10^{3}$ & 0.52 & 64.4 \\
24 & $4.2\times10^{-6}$ & $6.7\times10^{3}$ & $1.4\times10^{4}$ &
$5.5\times10^{3}$ & 0.51 & 64.1 \\
25 & $3.5\times10^{-6}$ & $7.9\times10^{3}$ & $1.2\times10^{4}$ &
$2.7\times10^{3}$ & 0.49 & 60.3 \\
\textbf{26} & $2.5\times10^{-6}$ & $5.0\times10^{3}$ & $7.5\times10^{3}$ &
$2.7\times10^{3}$ & 0.45 & 73.2 \\
\textbf{27} & $3.1\times10^{-6}$ & $5.4\times10^{3}$ & $1.0\times10^{4}$ &
$3.8\times10^{3}$ & 0.50 & 68.9 \\
\textbf{28} & $2.4\times10^{-6}$ & $4.3\times10^{3}$ & $6.4\times10^{3}$ &
$3.8\times10^{3}$ & 0.40 & 79.5 \\
\textbf{29} & $3.0\times10^{-6}$ & $4.6\times10^{3}$ & $9.7\times10^{3}$ &
$5.6\times10^{3}$ & 0.45 & 73.6 \\
\textbf{30} & $2.4\times10^{-6}$ & $3.8\times10^{3}$ & $5.5\times10^{3}$ &
$5.6\times10^{3}$ & 0.33 & 84.6 \\
\textbf{42} & $2.2\times10^{-6}$ & $3.6\times10^{3}$ & $3.6\times10^{3}$ &
$5.6\times10^{3}$ & 0.25 & 91.1 \\
\hline
\end{tabular}
\begin{list}{}{}
\item[$^\mathrm{a}$]The best models (see \S\ref{sec:results}) are indicated by
bold face.
\end{list}
\end{table*}

\begin{figure}
\resizebox{\hsize}{!}{\includegraphics{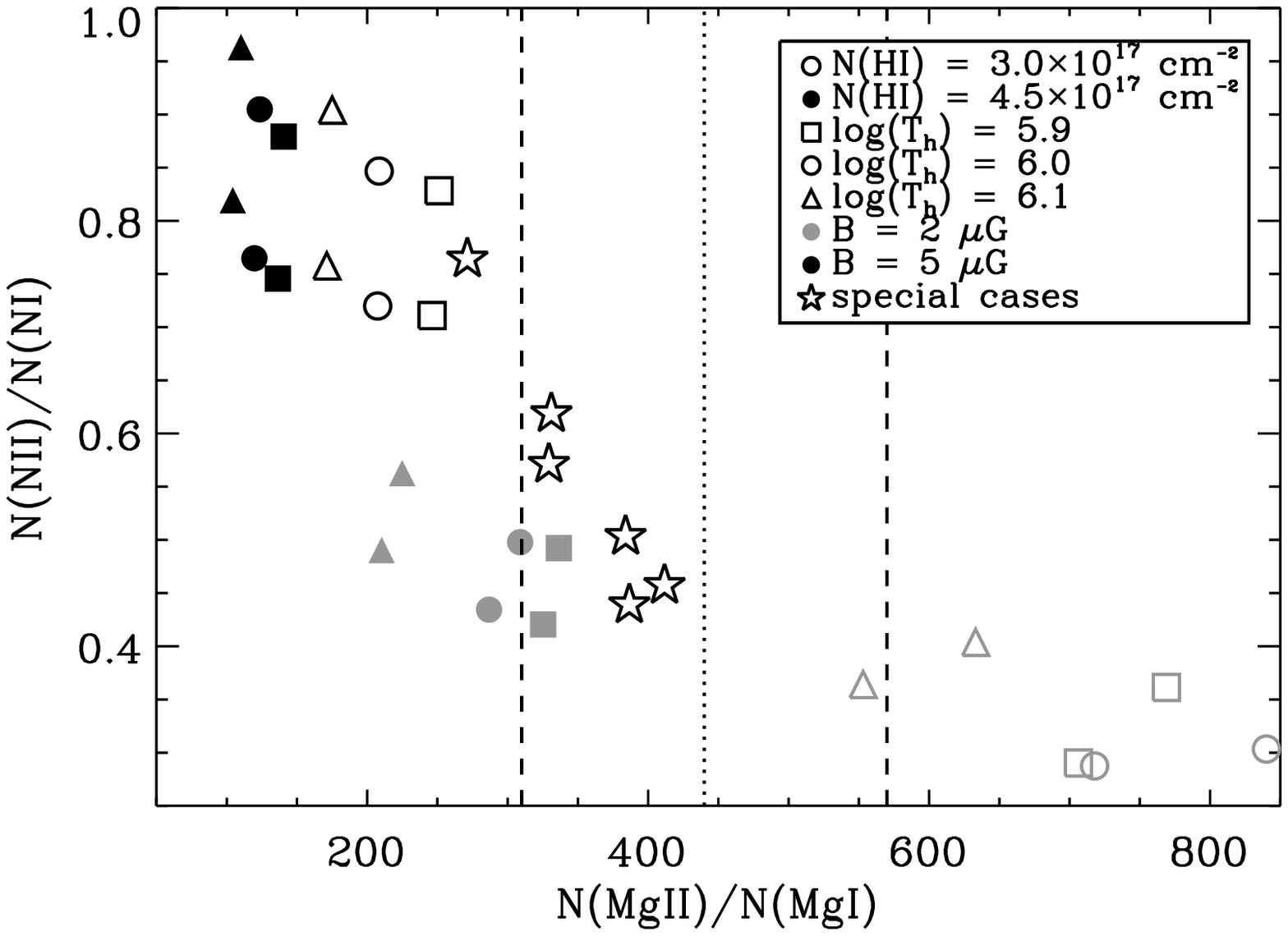}}
\caption{Model results for \NNII/\NNI\ versus \MgII/\MgI. The symbols have the
same meaning as in Figure \ref{fig:nHeT}.  In this case the models with higher
$n(\mathrm{H}^0)$ lie below and slightly to the left of those with lower
density.  \NNII/\NNI\ is an indicator of cloud ionization fraction while
\MgII/\MgI\ goes as $1/n_e$.  The dotted line is the observed value for
\MgII/\MgI\ and the dashed lines indicate the $1-\sigma$ error range for the
value. We see that models that match the observed ratio all correspond to
relatively low ionization, $X(\mathrm{H}) \sim 0.20 - 0.27$ in the CHISM.}
\label{fig:NMg}
\end{figure}

In Figure \ref{fig:nHeT} we show \nHeO\ and temperature of the CHISM for our
model calculations. In Figure \ref{fig:NMg} we show \NII/\NI\ vs.\ \MgII/\MgI,
illustrating an anti-correlation of the ratios caused by the fact that
\MgII/\MgI\ decreases with electron density while \NII/\NI\ indicates the
ionization level in the cloud.  The ionization of the CHISM is listed in Table
\ref{tab:ioniz} for Model 26, where commonly observed elements are listed.  The
abundances of He, Ne, Na, Al, P, Ar, and Ca were assumed, based on solar
abundances, and were not adjusted in the modeling process.  The abundances of
C, N, O, Mg, Si, S and Fe were adjusted for each model to match observed column
densities towards $\epsilon$ CMa (see \S \ref{sec:constraint}).  The elemental
abundances that have been assumed, with the exception of that for He, are not
expected to have any significant impact on the model results.

\begin{table}
\caption{Model 26 Results for Ionization Fractions$^\mathrm{a}$
\label{tab:ioniz}}
\centering
\begin{tabular}{llllll}
\hline\hline
Element & PPM & I & II & III & IV \\
\hline
H  & $10^{6}$ & 0.776    & 0.224 & --       & -- \\
He & $10^{5}$ & 0.611    & 0.385 & 4.36(-3) & -- \\
C  & 661      & 2.68(-4) & 0.975 & 0.0244   & 0.000 \\
N  & 46.8     & 0.720    & 0.280 & 8.52(-5) & 0.000 \\
O  & 331      & 0.814    & 0.186 & 4.71(-5) & 0.000 \\
Ne & 123      & 0.196    & 0.652 & 0.152    & 2.79(-6) \\
Na & 2.04     & 1.47(-3) & 0.843 & 0.155    & 6.34(-6) \\
Mg & 6.61     & 1.98(-3) & 0.850 & 0.148    & 0.000 \\
Al & 0.0794   & 5.37(-5) & 0.976 & 0.0118   & 0.0123 \\
Si & 8.13     & 4.21(-5) & 0.999 & 8.02(-4) & 3.10(-5) \\
P  & 0.219    & 1.35(-4) & 0.977 & 0.0232   & 9.29(-5) \\
S  & 15.8     & 6.47(-5) & 0.971 & 0.0288   & 1.95(-6) \\
Ar & 2.82     & 0.263    & 0.500 & 0.238    & 2.83(-6) \\
Ca & 4.07(-4) & 9.21(-6) & 0.0155 & 0.984   & 1.87(-4) \\
Fe & 2.51     & 7.01(-5) & 0.975 & 0.0245   & 5.75(-6) \\
\hline
\end{tabular}
\begin{list}{}{}
\item[$^\mathrm{a}$]Numbers less than $10^{-3}$ are written as x(y) where y
is the exponent and x is the mantissa (or significand).
\end{list}
\end{table}

\section{Discussion}

\subsection{Heliosphere Boundary Conditions}

As discussed above, the best models of those we calculated are determined by
the match to the CHISM \HeO\ density and temperature found by the \emph{in
situ} Ulysses measurements (\S \ref{sec:insitu}), combined with the matching
the LIC component column density ratios \MgII/\MgI\ and \CII/\CIIstar.  These
models span a fairly large range in the model parameters: $n_\mathrm{H} =
0.213 - 0.232$ \cc, $\log T_h = 5.9 - 6.1$ K, $B_0 = 0.05 - 3.8\, \mu$G, and
\NHI\ $= 3.0\times10^{17} - 4.5\times10^{17}$ \cmtwo.  Despite this, the
predicted values for neutral H density and electron density in the CHISM lie
within a remarkably small range: \nHO\ $= 0.19 - 0.20$ \cc, \nel\ $= 0.05 -
0.08$ \cc, for models that include the conductive boundary.  For the one model
without evaporation that is consistent with the data we find \nHO\ $= 0.186$
\cc, \nel\ $= 0.052$ \cc. Including the PUI Ne data as a constraint narrows
the density results to \nHO\ $\approx 0.19$ \cc\ and \nel\ $= 0.06 - 0.07$
\cc.  For these densities to stray out of this range would appear to require
significant errors in the underlying comparison data, e.g. \nHeO\ in the
CHISM, or substantial non-equilibrium ionization effects in the LIC.  The
variation in the interstellar radiation field between the different models
that match the data give us some degree of confidence that these densities are
not highly sensitive to the details of the radiation field.  The range \nel\
$= 0.05 - 0.08$ \cc\ corresponds to an electron plasma frequency of 2.0--2.5
kHz, which is the frequency of the mysterious weak radio emission detected
beyond the termination shock in the outer heliosphere
\citep{Gurnett:2005,MitchellCairnsetal:2004}.

\subsection{Hydrogen Filtration Factor}

Tracers of \HO\ inside of the termination shock, after filtration, include the
\HI\ \lya\ backscattered radiation, H pickup ions, and the slowdown of the
solar wind at distances beyond 5 AU from mass-loading by H PUIs.  The range of
\nHO\ found above to best fit the combined heliospheric \nHeO\ and LIC data
towards $\epsilon$~CMa, \nHO\ $ = 0.19 - 0.20$ \cc, represents the density of
neutral interstellar H atoms outside of the heliosphere, and removed from
heliospheric influences.  The hydrogen filtration factor, $f_\mathrm{H}$, can
be obtained from comparisons between these models and interstellar \HO\
densities at the termination shock as inferred from in situ observations of
interstellar H inside of the heliosphere.

The accompanying papers in this special section provide estimates of the
interstellar \HO\ density at the termination shock.  The solar wind slows down
due to massloading by interstellar H, yielding \nHO $=0.09 \pm 0.01$ \cc\ at
the termination shock \citep{Richardsonetal:2007}.  The density of H pickup
ions observed by Ulysses is inferred at the termination shock, yielding \nHO
$=0.11 \pm 0.01$ \cc; models of H atoms traversing the heliosheath regions
then yield for the CHISM \nHO $= 0.20 \pm 0.02$ \cc\ and \np $=0.04 \pm 0.02$
\cc, or \nel $\sim 0.05$ \cc\ \citep{Bzowskietal:2007}.  The radial variation
in the response of the interplanetary \lya\ 1215 \AA\ backscattered radiation
to the solar rotational modulation of the \lya\ ``beam'' that excites the
florescence yields \nHO $\sim 0.085 - 0.095$ \cc, depending on the heliosphere
model \citep{Pryoretal:2007}.  From these \nHO\ values at the termination
shock, we estimate that 43\%--58\% of the H-atoms successfully traverse the
heliosheath region, or $f_\mathrm{H} \sim 0.43-0.58$.
\citet{Mueller_etal_2007} evaluate filtration using five different
plasma-neutral models, and find a range of $f_\mathrm{H} = 0.52 - 0.74 $.  A
hydrogen filtration of $f_\mathrm{H} = 0.55 \pm 0.03$ is consistent with both
in situ data and radiative transfer models.

\subsection{Gas-Phase Abundances \label{sec:abundances}}

The LIC photoionization models are forced to match the observed set of
column densities (Tables \ref{tab:obs} and \ref{sec:photmod}).  The gas-phase
abundances of most elements are treated as free parameters that can be varied
in order to match observed column densities, so that the successful models
yield elemental abundances for the LIC that are automatically corrected for
unobserved H$^+$ (Table \ref{tab:abund}).  The exceptions are that He, Ne, and
Ar abundances, being unconstrained by the observations toward \eCMa, are
not adjusted but are assumed to be $10^5$ ppm, 123 ppm and 2.82 ppm,
respectively.  In the models, \NCIIstar\ is a constraint on both the C
abundance (in place of the heavily saturated \CII\ 1335\AA\ line) and \nel,
such that the product of the abundance and \nel\ is more tightly limited than
either quantity individually.  The requirement to match both \NCIIstar\ and
\nHeO\ effectively restricts the ionization fraction of H, which in turn
limits O and N ionization whose ionization fractions are tied by
charge-transfer to the H ionization at LIC temperatures.

\begin{table}
\caption{Elemental Gas Phase Abundances (ppm)\label{tab:abund}} 
\centering
\begin{tabular}{crrrrrrr} 
\hline\hline
 & \multicolumn{7}{c}{Element} \\ 
\cline{2-8} \\
Model No. & C & N & O & Mg & Si & S & Fe \\ 
\hline
14 & 589 &  40.7 & 295 & 5.89 &  7.24 &  14.1 & 2.24 \\
21 & 955 &  60.3 & 447 & 9.77 &  11.5 &  22.9 & 3.55 \\
25 & 631 &  66.1 & 437 & 7.76 &  10.0 &  19.5 & 3.09 \\
26 & 661 &  46.8 & 331 & 6.61 &  8.13 &  15.8 & 2.51 \\
27 & 759 &  64.6 & 437 & 8.71 &  10.7 &  20.9 & 3.31 \\
28 & 708 &  45.7 & 331 & 7.08 &  8.32 &  16.6 & 2.57 \\
29 & 813 &  64.6 & 437 & 9.33 &  11.0 &  21.9 & 3.39 \\
30 & 741 &  46.8 & 331 & 7.41 &  8.51 &  17.0 & 2.63 \\
42 & 724 &  39.8 & 295 & 6.76 &  7.76 &  15.1 & 2.34 \\
\hline
\end{tabular}
\end{table}

Early studies showed that the abundances of refractory elements in the very
local ISM are enhanced compared to abundances in cold disk gas
\citep{MarschallHobbs:1972,Stokes:1978,Frisch:1981}.  Throughout warm and cold
disk gas, the underabundances of refractory elements compared to solar
abundances (by factors of $10^{-1} - 10^{-4}$) are taken to represent
depletion onto interstellar dust grains \citep[e.g.][]{SS96}.  This view is
supported by the correlation found between elemental depletions and the
temperature characteristic of condensation at solar pressure and composition
\citep{Ebel:2000}, and assumes that there is a reference abundance pattern
that characterizes the cloud, and remains constant over the cloud lifetime as
atoms are exchanged between the gas and dust phases.  Below (\S
\ref{sec:gasdust}) we compare solar abundances with observed gas-phase
abundances to predict the gas-to-dust mass ratios for the LIC, based on the
assumption that LIC gas and dust have remained coupled over the cloud
lifetime.

An important question is whether the LIC has solar abundances.  Isotopes of
$^{18}$O and $^{22}$Ne isotopes measured in the ACR population suggest that
this is so.  ACRs are characterized by a rising particle flux for energies
below 10--50 MeV/nucleon, and this characteristic spectral signature is seen
for $^{16}$O, $^{18}$O, $^{20}$Ne, and $^{22}$Ne.  Ratios of
$^{16}\mathrm{O}/^{18}\mathrm{O} \sim 500$ and
$^{20}\mathrm{Ne}/^{22}\mathrm{Ne} \sim 13.7$ are found for both the ACRs and
solar material, indicating that the CHISM and solar material have similar
compositions \citep{Leskeetal:2000,Leske:2000}.  We therefore adopt solar
abundances as the underlying reference abundance pattern for the LIC.

Unfortunately a prominent uncertainty exists in the correct solar abundances of
volatile elements such as O and S, which have low condensation temperatures
($T_\mathrm{cond}$, 180 K and 700 K respectively), and noble elements such
as Ne and Ar.  Solar abundances are
determined from photospheric data (C, N, O, Mg, Si, S, Fe), the solar wind
(Ar, Ne), solar active regions (Ne), and helioseismic data (He); abundances of
non-volatile elements are also found from meteoritic data
\citep{GrevesseSauval:1998,Holweger:2001,Lodders:2003,GrevesseAsplund:2007}.
Solar abundances from these studies
are listed in Table \ref{tab:solarabundance}.  Our results, discussed below,
indicate that if the LIC has a solar abundance composition, as indicated by
the $^{18}$O and $^{22}$Ne data, then the lower abundances found by
\citet{GrevesseAsplund:2007} are preferred by our models.

\begin{table}
\caption{Solar Abundances (ppm) \label{tab:solarabundance}}
\centering
\begin{tabular}{lllll} 
\hline\hline
  &  Grevesse  &  Holweger  & Lodders   & Grevesse\\ 
  &  Sauval &  & & et al. \\ 
  &  (1998)         & (2001) & (2003)     & (2007)$^{a}$  \\ 
\hline
C  & $334 \pm 46 $ & $ 391 ({+110},{-86})$ & $290\pm 27$   & $275 \pm  34$ \\ 
N  & $ 84 \pm 12 $ & $85.3 ^{+25} _{-19}$ & $ 82\pm20$    & $67  \pm  10$ \\ 
O  & $683 \pm 94$  & $545^{+107} _{-90}$  & $579\pm66$    & $513 \pm  63$ \\ 
Ne & $121 \pm 17$  & $100 ^{+17} _{-15}$ & $ 91\pm21$    & $ 77 \pm  12$ \\ 
Mg & $ 38 \pm  4$  & $34.5 ^{+5.1} _{-4.5} $ & $ 41\pm 2$    & $ 38 \pm   9$ \\
Si & $ 35 \pm  4$  & $34.4^{+4.1} _{-3.7}$   & $ 41\pm 2$    & $ 36  \pm  4$ \\
S  & $ 22 \pm  5$  & - & $ 18\pm 2$    & $ 15 \pm   2$ \\ 
Ar & $2.54\pm 0.35$ & - & $4.24\pm 0.77$ & $1.70\pm 0.34$ \\ 
Fe & $ 32 \pm  4$   & $28.1^{5.8} _{4.8}$  & $ 35\pm 2$    & $ 32 \pm   4$ \\
\hline
\end{tabular}
\begin{list}{}{}
\item[$^\mathrm{a}$]Protosolar abundances are obtained by increasing the
photospheric abundances by 0.05 dex for elements heavier than He, as suggested
by \citet{GrevesseAsplund:2007}
\end{list}
\end{table}

\textbf{Ne:} In these models we have assumed the Ne abundance is 123 ppm
\citep{AndersGrevesse:1989}, which is based on a combination of photospheric
and interstellar data.  Solar system Ne abundances are difficult to measure
because of FIP effects, however values include 77 ppm \citep[][after adding
0.05 dex to account for gravitational settling of the
elements]{GrevesseAsplund:2007}, and $\sim 41$ ppm for solar wind in coronal
holes \citep{GloecklerGeiss:2007}.  The predicted densities of 
$n(\mathrm{Ne}^0)$ in the CHISM for models 26 and 28 (Table \ref{tab:sunres})
are in agreement with the most recent PUI results for Ne (Table
\ref{tab:obs}), and Ne densities as low as $\sim 100$ ppm are allowed when
filtration is included. 

The CHISM Ne abundances indicated by these results appear to be 
consistent with Ne abundances and ionization levels in the global ISM.  
The Ne abundance in the Orion nebula is 100 ppm \citep{Simpsonetal:2004}. 
\citet{TakeiNeon:2002} measured X-ray absorption edges formed by Ne and O 
in the interstellar gas and
dust towards Cyg X-2, and found abundances of Ne/H $\sim 92$ ppm and O/H $\sim
579$ ppm when both atomic and compound forms in the sightline were included.
\citet{JuettNeonIII:2006} observed the X-ray absorption edges of Ne and O
towards nine X-ray binaries which sampled both neutral and ionized warm
material, and found that the ionized states formed in the ionized material
have the ratio \NeIII/\NeII\ $\sim 0.23$.  This value is identical to the
predictions of model 26, \NeIII/\NeII\ $\sim 0.23$, a fortuitous agreement
that may indicate that the EUV radiation field in the CHISM is similar to the
generic galactic EUV field in the solar vicinity.

\textbf{Ar:} Solar Ar abundance determinations range between Ar/H $ \sim
1.4-5.0$ ppm (Table \ref{tab:solarabundance}); we have assumed Ar/H $= 2.82$
ppm.  The predicted Ar density at the Sun is within the uncertainties of the
PUI data, although the range of possible filtration factors (0.64 -- 0.95, see
\S\ref{sec:insitu}) also allow considerable leeway.

\textbf{O:} In the warm ISM such as the LIC, the ionization of oxygen and
hydrogen is tightly coupled over timescales of $\sim$100 years by charge
transfer \citep{FieldSteigman:1971}, so that the assumed \NHI\ combined with
$N$(\OI) measurements act to constrain the deduced O abundance in the gas.
The two best models (26, 28) correspond to O/H $= 331$ ppm, however the O
column density measurements are based on the saturated 1302\AA\ line, and have
$\sim 35$\% uncertainty.  The modeled LIC value of O/H $ \sim 331$ ppm
indicates that $\sim 35$\% of the O atoms are depleted onto dust grains.  An
oxygen filtration factor of $f_\mathrm{O} \sim 0.75 $ is required by the PUI
data and Model 26.

These models yield gas-phase O abundances that are consistent with
observations of more distant interstellar sightlines.
The ratio $N$(\OI)/$N$(\HI) is measured in both low and high extinction
clouds. \citet{Oliveiraetal:2003} used unsaturated \OI\ lines in the
910--1100\AA\ interval and found \OI/\HI\ $= 317\pm19$ ppm for $\sim30$
sightlines that included both types of material.  Sightlines with detected
\HH, \NH~$> 10^{20.5}$ \cmtwo\ and \meanH\ $= 0.1-3.3$ \cc, yield O/H $= 319
\pm 14$ ppm \citep{MeyerCardelliSofia:1997}.  A survey of 19 stars with an
average distance of 2.6 kpc by \citet{Andreetal:2003} found \OI/\HI\
$=408\pm14$ ppm, where the long sightline and high average value \NH/\ebv\
$=6.3 \times 10^{21}$ \cmtwo\ mag$^{-1}$ indicate a bias towards sightlines
containing many clouds that individually have low extinctions.  A larger
sample of 56 sightlines for a range of extinctions and distances show that
sightlines with higher average mean densities, \meanH, show O/H $=284 \pm 12$
ppm, versus O/H $=390 \pm 10$ ppm for stars with low values of \meanH\
\citep{Cartledge:2004}.  For comparison, solar abundance studies yield a range
of $\sim 450 -780$ ppm.

\textbf{C:} The best Models 26 and 28 yield a gas-phase abundance of C of C/H
$=661$ and 708 compared to solar abundances of $\sim 240 -500$ ppm,
which is consistent with our earlier results \citep{Slavin+Frisch_2006} 
indicating an overabundance of C in the LIC.  We speculate that shock
destruction of carbonaceous grains, perhaps combined with some local spatial
decoupling between carbonaceous and silicate grains, may explain these
findings.  

Singly-ionized carbon is an important coolant in the LIC (\S
\ref{sec:heatcool}), so the C overabundance is required to maintain the
temperature of the CHISM at the observed value.  The carbon abundance obtained
here indirectly depends on the \MgII$\rightarrow$\MgI\ dielectronic
recombination coefficient that determines the ratio \MgII/\MgI, since that
ratio is used as a criteria for the best models.  The same ionization
correction that gives the C abundance also successfully predicts Ne ionization
in global WPIM and the S abundances in the LIC, although this may be a
fortuitous coincidence.  In the adjacent sightline towards Sirius the LIC has
\NCII/\NHI\ $= 1,050$ ppm \citep{Hebrardetal:1999}.  An ionization correction
of 300\% is required to make this value consistent with solar abundances, and
such a large ionization correction is not consistent with the ionization
levels of $X(\mathrm{H}) \sim 20 - 26$\% found here.  In contrast, sightlines
with cold ISM show C abundances on the order of $ 135 \pm 46$ ppm
\citep{Sofiaetal:1997,Sembachetal:2000}.

\textbf{N:} The best models (26 and 28) find N/H = 46--47 ppm, compared to
solar values of $\sim 57 -110$ ppm.  These
results are consistent with the PUI results, N/H $\sim 19 - 47$ ppm, after
filtration factor uncertainties are included.  The N and O results
and favor an ISM abundance pattern for volatiles similar to 
the \citet{GrevesseAsplund:2007} photospheric abundances.

\textbf{S:} The best models predict S/H = 16--17, compared to solar
values of $\sim 13 -27$ (including uncertainties, see Table
\ref{tab:solarabundance}).  Sulfur is found to have little or no depletion 
onto dust grains in warm diffuse ISM \citep[e.g.][]{Wea99}.

\textbf{ Mg, Si, Fe:} These refractory elements are observed in the LIC gas
with abundances far below solar (factors of 3--15).  Approximately 92\%, 82\%,
and 77\% of the Fe, Mg, and Si, respectively, are presumably depleted onto
interstellar dust grains.  If \citep{GrevesseAsplund:2007} abundances are
assumed for the LIC, then to within the uncertainties the LIC dust has the
relative composition of Fe:Mg:Si:O = 1:1:1:4, as is consistent with amorphous
olivines MgFeSiO$_4$.  Fe and Si are dominantly singly ionized, while Mg has a
significant fraction ($\sim 15$\%) that is twice ionized.  The gas-phase
abundances of these refractory elements are highly subsolar, even after
ionization corrections are made, indicating that these elements are
substantially depleted onto interstellar dust grains.  In contrast to C,
however, the silicate dust in the LIC that carries the missing Mg, Si, and Fe
has experienced far less destruction than the carbonaceous grains.  

\textbf{\CaII, \NaI:} Weak lines of the trace ionization species \CaII\ and
\NaI\ are common diagnostics of ionization and abundance for interstellar
clouds, including the partially ionized LIC; \NaI\ is also frequently used as
a diagnostic of the H column density.  We note that our models show that the
ratios $N$(\CaII)/$N$(\NaI), $N$(\NaI)/$N$(H), and $N$(\NaI)/$N$(\HI) vary by
30\%, 77\%, and 93\%, respectively, between the best models (Models 26--30,
42).  As trace ionization species, the densities of \NaI\ and \CaII\
are highly sensitive to volume density, $n(\mathrm{Na}^0)$, $n(\mathrm{Ca}^+)
\propto n(\mathrm{H}) n_\mathrm{e}$.  We therefore conclude that \NaI\ and
\CaII\ are imprecise diagnostics of ionization levels, H density, and
abundances in warm partially ionized clouds.

\subsection{Gas-to-Dust Mass Ratio \label{sec:gasdust}}

Because the abundances are automatically corrected for unobserved H$^+$, we
use the model results to infer the total mass of the interstellar dust,
providing that the gas and dust in the LIC form a coupled and closed system
that evolves together as the cloud moves through the LSR.  The LIC LSR
velocity is 16--21 pc/Myr, so that a LIC origin related to the Loop I or
Scorpius-Centaurus superbubble would require that the LIC gas and dust
remained a closed system over timescales of 4--5 Myr
\citep{FrischSlavin:2006book,Frisch:1981}.  Gas-to-dust mass ratios calculated
from the best models (26 and 28) using the missing-mass argument\footnote{This
argument assumes that the ISM reference abundances, in this case solar
abundances, represent the sum of the atoms in the gas plus the dust
\citep{Frischetal:1999}.} are in the range \rgd$ = 149-217$, depending on
solar abundances.  The detailed information about \rgd\ for different
assumptions and the different models is listed in table \ref{tab:rgd}.

For comparison, \rgd\ determined from comparisons of \emph{in situ}
observations of interstellar dust inside of the solar system, compared
to the gas densities of these models, yield \rgd\ = 115--125
\citep[Table \ref{tab:rgd},][]{Landgraf_etal_2000,Altobelli_etal_2004}.
The in situ \rgd\ is an upper limit, since the smallest interstellar
dust grains (radii $ \le 0.15 ~\mu$m) with large charge-to-mass ratios (and
thus small Larmor radii) are excluded from the heliosphere by the interstellar
magnetic field which is draped over the heliosphere.  

For all the models, the \rgd\ determined from comparing in situ dust
measurements with the CHISM gas mass flux is lower than that determined by
assuming solar abundances and using the gas phase abundances we determine to
find \rgd.  This suggests that somehow the dust flowing into the heliosphere
is concentrated relative to the gas, compared to the overall LIC sightline
towards $\epsilon$ CMa.  The lower solar abundances of
\citet{GrevesseAsplund:2007} result in lower required depletions, and produces
stronger disagreements with \rgd\ determined from in situ data.  We do not
understand this result, which we have found previously
\citep{Frischetal:1999}.  Since \rgd\ is sensitive to the mass of Fe in the
dust grains \citep{Frisch+Slavin_2003}, we suggest that this difference may
indicate inhomogeneous mixing of the gas and silicate dust over the $\sim
0.64$ pc extent of the LIC.

\begin{table}
\caption{Gas-to-Dust Mass Ratios from Models and In Situ Observations
\label{tab:rgd}}
\centering
\begin{tabular}{l c ccccc c}
\hline\hline
& \multicolumn{7}{c}{Model$^\mathrm{a}$ } \\
\cline{2-8} \\
Source & 14 & 26 & 27 & 28 & 29 & 30 & 42 \\ 
\hline
GS98                 & 137 & 149 & 196 & 149 & 197 & 150 & 138 \\
Lodders              & 158 & 174 & 238 & 174 & 239 & 175 & 160 \\
Grevesse             & 194 & 217 & 321 & 217 & 323 & 218 & 196 \\
In Situ  	     & 115 & 116 & 123 & 116 & 125 & 116 & 107 \\
\hline \\
\end{tabular}
\begin{list}{}{}
\item[$^\mathrm{a}$]The source of the comparison solar abundances is listed in
column 1 (Table \ref{tab:solarabundance}.  The in situ dust flux is from
\citet{Landgraf_etal_2000}, corrected downwards by 20\% as recommended by
\citet{Altobelli_etal_2004} to account for side-wall impacts.
\end{list}
\end{table}

\subsection{Heating and Cooling Rates \label{sec:heatcool}}

\begin{table}
\caption{Major Heat Sources in LIC Gas$^\mathrm{a}$}
\label{tab:heating}
\centering
\begin{tabular}{lc}
\hline\hline
Source$^\mathrm{b}$ & Fraction of Heating \\
\hline
\ion{H}{i} & 0.657 \\
\ion{He}{i} & 0.248 \\
dust & 0.055 \\
\ion{He}{ii} & 0.016 \\
cosmic rays & 0.010 \\
\hline
\end{tabular}
\begin{list}{}{}
\item[$^\mathrm{a}$]Results for model 26.  Other models are qualitatively
the same, though there are some quantitative variations.
\item[$^\mathrm{b}$]For lines with ion names, the source here denotes the
ion that is photoionized.  Dust heating comes from photoelectric ejection by
photons of the background FUV radiation field.  Cosmic ray heating comes
from electron impact ionization of the gas and direct heating of the
electrons in the LIC plasma by the cosmic ray electrons.
\end{list}
\end{table}

The heating and cooling rates for Model 26 are listed in Tables
\ref{tab:heating} and \ref{tab:cooling}.  The primary heat sources are
photoelectrons from the ionization of \HO\ and \HeO, with dust and cosmic ray
heating contributing less than 7\% of the heating.  The dominant source of
cooling is the [\CII] 157.6 $\mu$m fine-structure line, making up 43\% of the
total.  This is more than twice the contribution of any other coolant.  Nearly
all the cooling is due to optical and infrared forbidden lines with many lines
contributing at the $\sim 1$\% level.  H recombination, free-free emission and
dust, through the capture of electrons onto grain surfaces, also contribute at
about a 2\% level. The importance of \CII\ as both a constraint on the C
abundance in our models as well as a major coolant means that any model that
aims to reduce the abundance of C to a solar level faces severe difficulties. 
The models with LIC temperatures in the $T_{\mathrm{He}^0} = 6\,300 \pm 340$ K
range indicated by the \emph{in situ} \HeO\ data all require supersolar
abundances of C.  The total heating/cooling rate for the LIC at the Sun for
this model is $3.55\times10^{-26}$ ergs \cc\ s$^{-1}$.

\begin{table}
\caption{Major Coolants in LIC Gas$^\mathrm{a}$}
\label{tab:cooling}
\centering
\begin{tabular}{lcc}
\hline\hline
Ion/Line & Fraction of Cooling \\
\hline
$[$\ion{C}{ii}$]\; 157.6\, \mu$m & 0.428 \\
$[$\ion{S}{ii}$]\; 6731\,$\AA\   & 0.145 \\
\ion{Fe}{ii} (total)           & 0.074 \\	
$[$\ion{Si}{ii}$]\; 34.8\,\mu$m  & 0.065 \\
$[$\ion{Ne}{ii}$]\; 12.8\,\mu$m  & 0.035 \\
$[$\ion{O}{i}$]\; 63.2\,\mu$m    & 0.028 \\
H recomb.\                     & 0.024 \\
dust                           & 0.024 \\
$[$\ion{Ne}{iii}$]\; 15.6\,\mu$m & 0.020 \\
$[$\ion{N}{ii}$]\; 6584\,$\AA\   & 0.018 \\
net free-free                  & 0.018 \\	
$[$\ion{O}{i}$]\; 6300\,$\AA\    & 0.017 \\
$[$\ion{O}{ii}$]\; 3727\,$\AA\   & 0.011 \\
$[$\ion{Ar}{ii}$]\; 6.98\,\mu$m  & 0.011 \\
\hline
\end{tabular}
\begin{list}{}{}
\item[$^\mathrm{a}$]Results for model 26.  Other models show similiar
results.
\end{list}
\end{table}

\subsection{Radiation Field\label{sec:disc-rf}}

Recently it has been proposed that a significant portion of the SXRB can be
attributed to charge-transfer (a.k.a.\ charge exchange) between the solar wind
ions (e.g., \OVII\ and \OVIII) and interstellar neutrals \citep{Cravens_2000,
Snowden_etal_2004,Wargelin_etal_2004,Smith_etal_2005,Koutroumpa_etal_2006}.
While it seems at present that some fraction of the low energy X-rays are from
this mechanism, it is unclear how large that fraction is. We note that basing
the properties of the local hot plasma in the galactic plane on SXRB emission
at energies $E > 0.3$ keV is problematical.  \citep{BellmVaillancourt:2005}
have compared the Wisconsin B and Be band data with the ROSAT R12 data, and
concluded that the observed anticorrelation between R12 and \NHI\ indicates
that more than 34\% of the SXRB generated in the Galactic disk must come from
the Local Bubble.  They also concluded that a heavily depleted plasma with log
$T \sim 5.8$ is consistent both with the
\citet{McCammon_etal_2002,Sanders_etal_2001} X-ray spectral data, and the
upper limits set on the EUV emission by CHIPS \citep{CHIPS:2004}.  When the
\citet{RobertsonCravens:2003} models of SXRB production by charge-transfer
with the solar wind are considered, then only half of the SXRB in the plane is
required to arise from a hot local plasma.  We also note that the atomic
physics for the calculation of the low energy part of the emission is still
quite uncertain (V.\ Kharchenko, private communication).  At this point we
take the simple approach of ignoring the charge-transfer emission, though we
plan to consider its possible impact in future work by reducing the assumed
SXRB flux from hot gas.  As noted previously, a lower SXRB flux due to a lower
pressure in the hot gas does not necessarily have any impact on our calculated
flux from the evaporative cloud boundary.

\subsection{LIC Pressure}
The strength of the interstellar magnetic field in the LIC is unknown, though
modeling of its effects on the heliosphere sets some constraints.  Our best
models (26 and 28) presented here have a thermal pressure of $\sim 2100$
\cc~K for the LIC.  If the thermal and magnetic pressures are equal, this
indicates a magnetic field strength $B \sim  2.7\, \mu$G, in agreement with
field strengths for these models. As noted is \S\ref{sec:radfield}, the main
effect of the field strength in the models is to regulate the pressure in the
evaporative cloud boundary, which in turn affects the flux of diffuse EUV
radiation incident on the cloud. The amount of EUV flux helps determine the
temperature in the cloud, which is how the observational constraints fix the
magnetic field strength in the context of our modeling.  Thus we do not
explicitly fix the magnetic field strength with the goal of achieving
equipartition and indeed some of our successful models have lower or higher
field strengths.  It is probably coincidental that the field strength required
to match the in situ \HeO\ temperature for our best models is also close to
the equipartition field strength, but it is at least encouraging that this
field strength is consistent with our photoionization models.  We note that if
thermal, cosmic ray, and magnetic pressures are approximately equal the LIC
has a pressure of $\sim 6300$ \cc\ K.

\subsection{Comparisons with other LISM Sightlines}
There have been a number of efforts to understand the LISM ionization and
abundances \citep{Frisch_etal_1986,Cheng+Bruhweiler_1990,Lallement+Bertin_1992,
Vallerga:1996,Lallement+Ferlet_1997,Holberg_etal_1999,Kimura_etal_2003}. 
The studies that attempt to derive gas phase elemental abundances find a range
of results, generally fairly consistent with ours.  A point of particular
interest is the abundance of carbon that is surprisingly overabundant in our
results. As an example, \citet{Kimura_etal_2003} find (based on four
sightlines and excluding the \eCMa\ and  $\beta$ CMa sightlines), a subsolar C
abundance in contradiction with our results.  Results for thirteen sightlines
from \citet{Redfield+Linsky_2004} with velocity components consistent with the
LIC velocity vector show that $N($\ion{C}{ii}$)/N($\ion{O}{i}$) > 1$ for 8 of
them, especially those at lower column density indicating a solar or
supersolar C abundance.  For our best models $N($\ion{C}{ii}$)/N($\ion{O}{i}$)
\approx 2$.  Our series of studies are unique in that we model the radiation
field incident on the cloud, include radiative transfer effects, and calculate
the thermal equilibrium within the cloud.  The ionization varies through the
cloud as does the temperature and density (slightly) and we compare
observations within the heliosphere with the physical conditions at that point
in the cloud rather than basing the model on line-of-sight averages.

Our present results indicate that $n$(\NII)/$n$(\NI)$\sim 0.32 - 0.50$ at the
solar location, with N becoming more ionized as the sightline approaches the
cloud surface.  The column density ratio is thus higher, ranging from $0.38 -
0.62$. Observed values of $N$(\NII)/$N$(\NI) toward other nearby stars
are $0.58^{+0.56}_{-0.77}$ toward Capella \citep{Wood_etal_2002},  $1.29 \pm
0.23$ toward HZ43 \citep{Kruk_etal_2002}, $1.91 ^{+0.87}_{-0.69}$ towards
WD1634-573 \citep{Lehner_etal_2003}, and $1.13 \pm 0.24$ towards $\eta$ UMa
(Frisch et al. in preparation).  The total \HI\ column density towards each of
these stars is greater than the \NHI$\sim 4 \times 10^{17}$ \cmtwo\ found for
the best models here.  The nearest of these stars, Capella, has an ionization
comparable to that of the LIC.  The two high-latitude stars HZ43 and $\eta$
UMa appear to sample low opacity regions where the ionization is larger than
at the Sun, as does the WD1634-573 sightline that appears to cross the nearby
diffuse \HII\ region seen towards $\lambda$ Sco \citep{York:1983}.  As we have
noted the $\epsilon$ CMa line of sight is special because that star is the
dominant source of stellar EUV photons for the LIC. Thus for sightlines at a
large angle from the $\epsilon$ CMa sightline, if the \HI\ column between
points along the sightline and $\epsilon$ CMa is small the apparently high
column points are subject to a strong EUV field.  Such geometry dependent
ionization effects can be important for non-spherical clouds subject to a
strongly spatially variable ionizing radiation source.

The variation in the fractional ionization of the CLIC gas has a direct
impact on our understanding of the distribution and physical properties of low
column density clouds for several reasons.  (1) Abundances of elements with
FIP $<13.6$ eV must always be calculated with respect to \NHI $+$ \NHII\ for
very low column density clouds.  (2) Cloud geometry affects the opacity of
observed sightlines so that the opacity to ionizing radiation is not directly
traced by the observed value of \NHI.  For lines of sight other than that
towards \eCMa, this could require more complex radiative transfer models in
which the difference between the line of sight toward the star and that toward 
one of the primary sources of ionizing flux, \eCMa, at each point is taken
into account.

\section{Conclusions}
There are many uncertainties regarding the detailed properties of the ionizing
interstellar radiation field incident on the LIC.  The data we have on the
LIC, both from absorption line studies and \emph{in situ} measurements by
spacecraft in the heliosphere, provide us with strong constraints on the
ionization and composition of the LIC and particularly the CHISM.  By
exploring a range of models for the ISRF we find that while a fairly broad
range of radiation fields can produce photoionization consistent with the
data, other outputs from the models fall within a relatively narrow range of
values.  Our results for the models explored in this paper in which we require
our models to be consistent with the LIC component of the absorption lines
observed towards \eCMa\ include:
\begin{enumerate}
\item For a range of assumptions regarding the \HI\ column density of the LIC,
\NHI\ $= (3.0 - 4.5)\times10^{17}$ \cmtwo, and temperature of the hot gas of
the Local Bubble, $\log T_h = 5.9, 6.0$ and 6.1, we are able to find model
parameters that allow a match of the model results with best observed
quantities, $n($\HeO$)$, $T($\HeO$)$ and \NMgII/\NMgI.  For these models we
assume that the cloud is evaporating because of thermal conduction between the
hot Local Bubble gas and the warm LIC gas and include the emission from the
cloud boundary.
\item For the best models in terms of fits to data, the required input
parameters are: initial (i.e.\ at the outer edge of the cloud) total H
density, $n(\mathrm{H}) \approx 0.21 - 0.23$ \cc; and cloud magnetic field,
$B_0 \leq 3.8\, \mu$G.
\item If we assume that the magnetic field configuration reduces thermal
conductivity at the boundary enough to prevent evaporation and ignore any
radiation from the cloud boundary, we find that for most cases the radiation
field does not cause sufficient heating to maintain the LIC at the temperature
observed, $T = 6,300\pm340$~K.  One set of parameter choices, though, yields
a successful model. These parameters are \NHI\ $= 4.5 \times 10^{17}$ \cmtwo\
and $\log T_h = 6.1$.
\item Despite the wide range of possible input parameters, the output values
for quantities important for shaping the heliosphere are confined to a fairly
small range: \nHO\ $= 0.19  - 0.20 $ \cc, and \nel\ $=0.05 - 0.08$ \cc.
\item A H filtration factor of  $f_\mathrm{H} = 0.55 \pm 0.03 $ yields good
agreement between the radiative transfer model predictions for \nHI\ in the
CHISM, and \nHI\ at the termination shock as found from observations of PUIs,
the \HI\ \lya\ glow, and the solar wind slow-down in the outer heliosphere.
This filtration value is also consistent with heliosphere models of the
ionization of interstellar H atoms traversing the heliosheath regions.
\item Elements with ionization potentials $13.6 - 25$ eV, e.g.\ H, He, N, 
O, Ne, and Ar, are partially ionized with ionization fractions of 
$\sim 0.2 - 0.7$.
\item By requiring that the models match the column densities derived from
absorption line data we are able to determine the necessary elemental
abundances for several elements.  We find that the abundances of N and O may
be somewhat sub-solar.  Sulfur is roughly solar, and C is substantially
super-solar. Mg, Si and Fe are all sub-solar by factors of $3-15$.  The
depletions of Fe, Mg, Si and O in the LIC are consistent with a dust
population consisting of amorphous silicate olivines MgFeSiO$_4$, though other
compositions for the dust are possible as well.  We conclude that any
carbonaceous dust in the LIC must have been destroyed, while silicate dust has
persisted.  Except for the gas-to-dust mass ratio, these results are in better
agreement with the lower solar abundances of \citet{GrevesseAsplund:2007}.
However we note that the O and Ne abundances of \citet{Lodders:2003} are in
better agreement with other astronomical data such as the X-ray absorption
edges.
\item The gas-to-dust mass ratio derived from missing mass in the gas-phase
for our best models depends strongly on the assumed reference abundance set
and range from $137- 323$.  Our two best models, nos.\ 26 and 28, give a range
of $149 - 217$.  For these same models \rgd\ $= 115 - 125$ based on the
observed flux of dust into the heliosphere.  The discrepancy of these
values is minimized, in fact leading to consistency within the errors, if one
assumes an abundance set such as that of GS98 which has large abundances of
the metals.  The GS98 abundances lead to substantial O depletion, however,
which is not easily explained and conflict with the S abundances found for
models 26 and 28.
\item These models also show that the densities of the trace ionization
species \CaII\ and \NaI\ are extremely sensitive to density and ionization. 
Therefore the ratios $N$(\CaII)/$N$(\NaI), $N$(\NaI)/$N$(\HI), and
$N$(\NaI)/$N$(H) are, by themselves, inadequate diagnostics of warm low density
diffuse gas.  
\end{enumerate}

\begin{acknowledgements}
We would like to thank George Gloeckler for sharing data with us prior
to publication, and Alan Cummings for pointing out that the ACR
isotopic data indicate that the LIC abundances are solar.  We also thank the
International Space Science Institute in Bern, Switzerland for hosting the
working group on ``Interstellar Hydrogen in the Heliosphere.''  This research
was supported by NASA Solar and Heliospheric Program grants NNG05GD36G and
NNG06GE33G to the University of Chicago, and by the NASA grant NNG05EC85C to
SWRI.
\end{acknowledgements}

\bibliography{merged}
\bibliographystyle{aa}
\end{document}